\documentclass[5p,times]{elsarticle}

\usepackage{booktabs} 
\usepackage{amssymb}
\usepackage{amsmath}
\usepackage{lineno,hyperref}
\modulolinenumbers[5]
\usepackage[htt]{hyphenat}

\journal{Future Generation Computer Systems}

\usepackage{caption}
\usepackage{subcaption}

\usepackage{color,xcolor}

\usepackage{listings}
\newcommand*\lstinputpath[1]{\lstset{inputpath=#1}}
\lstinputpath{listings}

\graphicspath{{figures/}}

\definecolor{pblue}{rgb}{0.13,0.13,1}
\definecolor{pgreen}{rgb}{0,0.5,0}
\definecolor{pred}{rgb}{0.9,0,0}
\definecolor{pgrey}{rgb}{0.46,0.45,0.48}
\definecolor{light-gray}{gray}{0.9}

\begin{document}

\begin{frontmatter}

\title{Exploiting Inherent Elasticity of Serverless in Irregular Algorithms\tnoteref{mytitlenote}}
\tnotetext[mytitlenote]{This article is an extension of a work published in IEEE CLOUD’20 (París et al., 2020)~\cite{elasticParis}. The major new contributions of this article are: i) Performance study with Betweenness Centrality algorithm ii) Algorithms characterization iii) Cost analysis of UTS optimization}


\author[urvaddress]{Gerard Finol\corref{mycorrespondingauthor}}
\cortext[mycorrespondingauthor]{Corresponding author}
\ead{gerard.finol@urv.cat}

\author[urvaddress]{Gerard París}
\ead{gerard.paris@alumni.urv.cat}

\author[urvaddress]{Pedro García-López}
\ead{pedro.garcia@urv.cat}

\author[urvaddress]{Marc Sánchez-Artigas}
\ead{marc.sanchez@urv.cat}

\address[urvaddress]{Universitat Rovira i Virgili, Tarragona, Spain}

\begin{abstract}
Serverless computing, in particular the Function-as-a-Service (FaaS) execution model, has recently shown to be effective for running large-scale computations, including MapReduce, linear algebra and machine learning jobs, to mention a few. Despite this wide array of applications, little attention has been paid to highly-parallel applications with \emph{unbalanced} and \emph{irregular} workloads. Typically, these workloads have been kept out of the cloud due to the impossibility of anticipating their computing resources ahead of time, frequently leading to severe resource over- and underprovisioning situations. Our main insight in this article is, however, that the \emph{elasticity} and \emph{ease of management} of serverless computing technology can be a key enabler for effectively running these problematic  workloads for the first time in the cloud. More concretely,  we demonstrate that with a simple serverless executor pool abstraction, a data scientist can achieve a better cost-performance trade-off than a Spark cluster of static size built upon large EC2 virtual machines (VMs). To support this conclusion, we evaluate three irregular algorithms: the Unbalanced Tree Search (UTS), the Mandelbrot Set using the Mariani-Silver algorithm, and the Betweenness Centrality on a random graph. For instance, our serverless implementation of UTS is able to outperform Spark by up to $55\%$ with the same cost. We also show that a serverless environment can outperform a large EC2 VM in the Betweenness Centrality algorithm by a $10\%$ using the same amount of virtual CPUs. This provides the first concrete evidence that highly-parallel, irregular workloads can be efficiently executed using purely stateless functions with almost zero burden on users \textemdash~i.e., no need for users to understand non-obvious system-level parameters and optimizations. Furthermore, we show that UTS can benefit from the FaaS' pay-as-you-go billing model, which makes it worth for the first time to enable certain application-level optimizations that can lead to significant improvements (e.g., of $41\%$)  with negligible increase in cost.
\end{abstract}


\end{frontmatter}


\section{Introduction}

Serverless computing has been considered the next natural step in the evolution of cloud computing. Features like elasticity, no need to provision servers, and pay-as-you-go billing, have attracted a lot of interest both in academy and industry.

Since mid-2010s, all public cloud providers have presented serverless computing offerings under the  Function-as-a-Service (FaaS) execution model. Services like AWS Lambda, Google Cloud Functions, IBM Cloud Functions or Azure Functions, allow developers to write modular pieces of code that can be executed in response to certain events. The FaaS model was initially conceived to execute short and event-driven stateless computations, so it imposes a set of architectural constraints and operational limits on memory and CPU resources, network connectivity and function concurrency.

Despite these constraints, researchers have stepped beyond the limits of serverless functions to build more complex applications~\cite{berkeleyServerless}. Simplicity and fine-grained accounting have attracted researchers to this new compute abstraction that brings new opportunities to better support changing workloads. So, serverless functions have been used to run data-parallel algorithms~\cite{PyWren2017,IBMPyWren, ripple}, video encoding~\cite{excamera}, large-scale linear algebra operations~\cite{numpywren} or stateful computations~\cite{statefulCrucial, cloudburst}, among other applications. 

Unfortunately, no attention has been paid to a class of problems that exhibit high levels of parallelism, but with \emph{unbalanced} and \emph{irregular} workloads.  Running these algorithms remains challenging for many users due to the number of decisions to make ahead of time to efficiently execute them, ranging from provisioning and cluster management to deep understanding of system-level parameters. For instance, an efficient execution of the Unbalanced Tree Search (UTS) benchmark~\cite{prins03} requires fine-tuning task load-balancing~\cite{lifelineBasedGLB},  especially at large scales~\cite{ppoppTardieuHCGKSSTV14},  compelling users to deal with complex programming models and optimizations. Indeed, the goal of most of these optimizations is to mask the overheads of dynamic thread creation in static multithreaded systems.  Provisioning a cluster of any static size will either slow down the job or leave the resources under-utilized.  With a serverless solution, however, users may obviate the need to explicitly configure and manage on-demand compute units (e.g., VMs)~\cite{berkeleyServerless, PyWren2017,IBMPyWren, numpywren}. 

While simplicity is a key feature of serverless computing, an active research question is to determine which workloads are a natural fit for serverless computing. Prior literature has proven that a FaaS execution model can deliver high performance in many tasks ~\cite{PyWren2017,IBMPyWren, numpywren, excamera, crucial}. Nevertheless, superior performance of serverless computing has been generally achieved at the expense of greater monetary costs, sometimes leading to a poor cost-performance trade-off. To wit,~\cite{berkeleyServerless} reports serverless solutions with up to $7$x increase in cost compared with a traditional solution with VMs. 

Our goal in this work is to evaluate if the inherent elasticity of serverless is ideally suited for the implementation of irregular and unbalanced algorithms. More concretely, our focus is to ascertain whether serverless computing can simplify the programming model by pushing out load balancing from it, while achieving a better cost-performance trade-off than traditional VM-based solutions, which do not have the flexibility required for such workloads. This article provides a positive answer to this question, making serverless computing a compelling approach to efficiently run unbalanced algorithms at large scale. Actually, we demonstrate in the article that a basic serverless executor abstraction, mimicking the Java concurrency library, but launching Java threads as cloud functions, is enough to deliver a better cost-performance ratio than VM-based solutions. Simply put, with this naive executor, there is no need for users and developers to deal with non-obvious system-level components such as task schedulers to achieve good performance. It is enough for them to abide by the well-known, \emph{thread pool pattern}~\cite{threadpool} to code their applications. For instance, our serverless implementation of UTS is able to outperform Spark by up to $55\%$ with the same cost. We are the first to provide concrete evidence that highly-parallel, irregular workloads can be efficiently executed using purely stateless functions with almost zero burden on users.

These excellent results start to pave the way towards a more transparent and elastic execution of algorithms at large scale. Coulouris's textbook~\cite{coulouris} defines transparency as \textit{the concealment from the user and the application programmer of the separation of components in a distributed system}, and also defines a specific form of transparency -- scaling transparency -- as the property that \textit{allows the system and applications to expand in scale without change to the system structure or the application algorithms}. Our ultimate aim is thus to provide an elastic programming model that maintains the simplicity of development of single-machine applications when scaling up to thousands of cloud cores~\cite{serverlessEndGame}.

\subsection{Contributions}

In order to assess if FaaS is an appropriate execution environment for unbalanced algorithms, we implement and evaluate three challenging algorithms that exhibit irregular task-parallelism: the Unbalanced Tree Search (UTS)~\cite{prins03}, the Mariani-Silver algorithm~\cite{munafo10}, and the Betweenness Centrality (BC) benchmark~\cite{ssca2}. The first two also present nested parallelism (i.e. parallel tasks that generate new parallel tasks) while the third can be statically partitioned.
UTS is a well-known benchmark that has been widely used to evaluate task parallelism and load balancing techniques in several parallel computing architectures and different programming models~\cite{uts, ppoppTardieuHCGKSSTV14, resilientx10}. The Mariani-Silver algorithm is a recursive optimization technique to compute the Mandelbrot set. This algorithm has been used as an interesting case study of dynamic parallelism in CUDA~\cite{nvidiaMandelbrot}. And Betweenness Centrality is a graph algorithm that computes a centrality metric for each vertex of a graph. The implementation used in this paper is taken from the fourth kernel in the SSCA2 (Scalable Synthetic Compact Application 2) v2.2 benchmark~\cite{ssca2}.

Remarkably, the serverless executor construct allows users to preserve the simplicity of the original algorithms,  while achieving an optimal trade-off between cost and performance. Traditionally, algorithms such as UTS require significant effort to scale out~\cite{ppoppTardieuHCGKSSTV14}. Consequently, providing a serverless solution that can rival with HPC-like solutions with close to zero configuration complexity, yet keeping the original simplicity of the algorithms is highly desirable.  We believe that this work is a good starting point for more sophisticated implementations yet to come.

To verify the suitability of serverless computing to efficiently run unbalanced algorithms at large scale, we compare our serverless implementation of UTS with an existing Spark implementation~\cite{resilientx10}. Also, we compare the cost-performance ratio of the serverless implementations of the algorithms against large VM instances. The comparison between cloud functions and VM instances provides novel insights into the parallel efficiency of services like AWS Lambda and large EC2 VMs, including low-level issues such as the effect of Intel’s HyperThreading on the efficiency of compute-intensive, unbalanced tasks.

In summary, our contributions are:
\begin{enumerate}
  \item We present a serverless hybrid executor middleware, based on the Java concurrency library that executes Java threads over serverless functions. Our elastic programming model provides scaling transparency by efficiently combining local and remote computing entities.
  
  \item We have implemented\footnote{Source code available at https://github.com/gfinol/elastic-exploration} three unbalanced and irregular task-parallel applications using the serverless executor: UTS, Mariani-Silver and Betweenness Centrality. We provide the first evidence that the FaaS execution model can be leveraged to efficiently run this kind of algorithms. 

  \item We analyze the cost of running such algorithms in serverless environments and identify those situations where a serverless model can be cost-efficient. As shown in our evaluation, a performance benefit between $20\%$ to $55\%$ can be attained without increasing monetary costs compared to a Spark cluster.

  \item We have characterized these irregular algorithms using the Coefficient of Variation, the task generation rate and the CDF of the execution time. We use this characterization to optimize the configuration and deployment of these algorithms to the serverless environment.
\end{enumerate}

Section \ref{sec:related} contains related work and background for this article. In section \ref{sec:executors} we describe the serverless executor used as well as our novel hybrid executor. Methodology of the validation is presented in the section \ref{sec:methodology}, in particular, the metrics used and the algorithms studied are explained along with their characterization. Section \ref{sec:performance} contains a performance analysis where we evaluate, inter alia, the effects of hyperthreading and shared resources, the UTS optimizations and the performances of the hybrid executor. Section \ref{sec:cost} contains a cost-performance study where we compare the serverless and hybrid executors with Spark clusters and large VMs. Finally, conclusions can be found in section \ref{sec:conclusion}.

\section{Background / Related work} 
\label{sec:related}

Traditionally, VM clusters and HPC systems have been used to deal with highly unbalanced algorithms. Having a fixed amount of allocated resources generates the need to maximize their utilization, so users do not end up paying for a large number of idle CPU cycles. This need has typically been met with the implementation of complex dynamic load balancing and job stealing techniques.

One example of unbalanced algorithm is the Unbalanced Tree Search. The UTS benchmark was initially proposed by Prins et al. \cite{prins03} in 2003 and has since been used as an irregular application for evaluating load-balancing algorithms in several parallel computing architectures and programming models, inlcuding MPI~\cite{DinanOSPST07}, Unified Parallel C~\cite{OlivierP08} or X10's APGAS~\cite{lifelineBasedGLB}. However, all these programming models are not elastic and have been specifically designed to perform load balancing. 

On the other hand, serverless platforms (e.g. AWS Lambda, Google Cloud Functions) were initially designed to run short-running, event-driven and stateless computations. In the recent years several execution frameworks~\cite{PyWren2017, lithops, excamera, ggfouladi19, numpywren, carreira2019cirrus, locus, crucial, graphless,kappa, wukongsocc2020} have been proposed to leverage the scalability of the FaaS model to build more complex parallel applications on top of serverless platforms. 

Others frameworks, like Mashup~\cite{mashup} propose a hybrid environment combining VM clusters with serverless computing. Mashup targets HPC and scientific workflows and focuses on the task scheduling. For a given directed acyclic graph (DAG) of tasks to perform, it optimizes whether the task should be run in a VM cluster or in cloud functions. In order to optimize the execution time, it automatically measures time duration of the task for both VM cluster and cloud functions. These experimental measures are ill-suited for dynamic algorithms, where task duration cannot be anticipated at al.

All these available frameworks have mainly been applied to embarrassingly parallel problems. Instead, irregular algorithms have received little attention, and are only considered in frameworks designed for specific applications, like graph processing (Graphless~\cite{graphless}) or large-scale linear algebra (numpywren~\cite{numpywren}). Both of them use a data driven approach. Graphless partitions the set of nodes of a graph and distributes them to cloud functions. This approach can be beneficial for algorithm like BC, where the computation of the centrality of a node is independent of the others nodes' centrality. Numpywren implements common algebra algorithms (e.g., matrix factorization) as a set of independent DAG tasks, where each task operates over a particular subset of the entire dataset. Unbalanced algortihms like UTS are not suitable to this kind of frameworks because we can not know beforehand the structure of the tree. Therefore, there is no way to find an optimal partition or correctly represent computations into a DAG of tasks. To the best of our knowledge, there is no previous work studying the feasibility of serverless for elastic and irregular CPU-intensive algorithms with high levels of parallelism.

\section{Serverless Executor Middleware}
\label{sec:executors}

For the validation in this work we have borrowed the \texttt{ServerlessExecutor} from Crucial \cite{statefulCrucial}. In addition, based on this executor, we have implemented a hybrid executor that combines a VM with serverless functions.

\subsection{Serverless Executor}
The \texttt{ServerlessExecutor}  from Crucial is a general-purpose serverless executor based on thread pool pattern from Java Concurrency library. The fragment of code that runs as a serverless function is written as a typical Java \texttt{Callable} task, so the developer can use remote cloud functions as if they were local threads. In particular, we have used the \texttt{ServerlessExecutor}, which processes submitted \texttt{Callable} tasks and runs them as serverless functions.

This serverless implementation follows a master-worker model to support dynamic parallelism of nested algorithms. The client application is responsible to invoke remote tasks as cloud threads and asynchronously retrieve results with a \textit{Future} abstraction. The client application has also the logic that decides to split tasks. Therefore, the client application collects return values from remote tasks through a local thread pool. These return values are inserted to a queue. A local thread keeps polling the queue and eventually splits pending tasks before invoking new remote threads. Serverless tasks are completely stateless.

FaaS platforms impose some limits on function concurrency that the client application must be aware of. For example, AWS Lambda has a default limit of $1,000$ concurrent function executions (a limit that can be increased upon request). Therefore, the local thread pool must be limited to avoid function throttling exceptions. The invocation frequency is also limited, with great differences in default values between providers: $5,000$ per minute in IBM Cloud Functions and $10,000$ per second in AWS Lambda. In this case, the invocation frequency limit of AWS Lambda is high enough to run our experiments.

To compare performance and cost of serverless functions against a local multithread version, we have implemented parallel multithread versions of the algorithms following the same master-worker approach. In fact, serverless and multithread implementations are very similar. Because we use Java Concurrency API, an existing Java multithread implementation is easily adaptable to serverless. 

\begin{figure}
  \centering
  \includegraphics[width=0.9\linewidth]{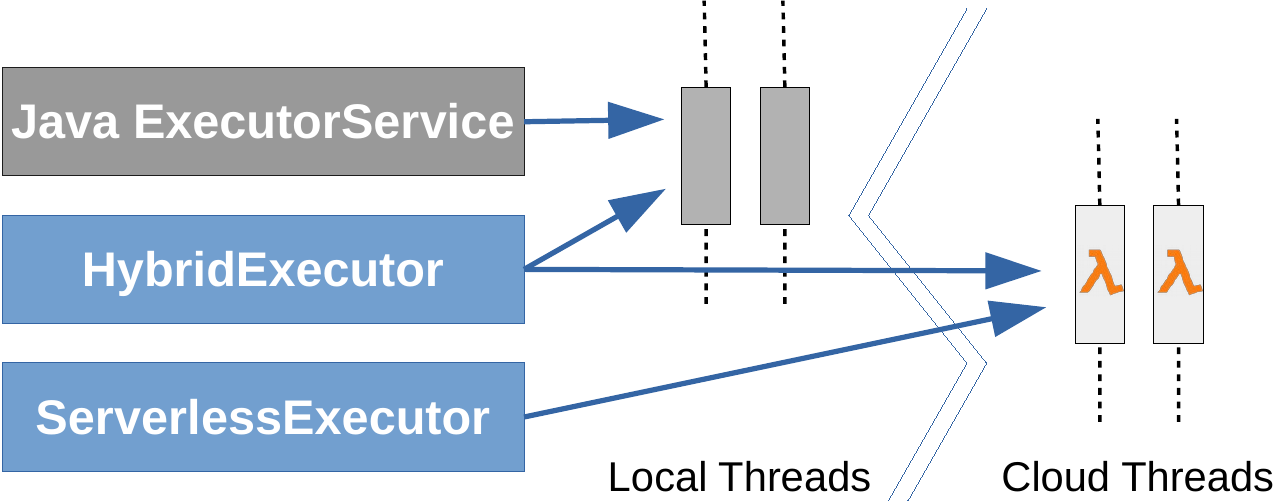}
  \caption{
    Serverless and Hybrid executors.
  }
  \label{fig:executors}
\end{figure}

\subsection{Hybrid Executor}
We have implemented a \texttt{HybridExecutor} that combines local threads from a VM and remote serverless functions (as depicted in Figure \ref{fig:executors}). This hybrid version is transparent from the application point-of-view: \texttt{Callable} tasks are submitted to the \texttt{HybridExecutor}, which decides whether to run them as local threads or remote serverless functions. Listing \ref{lst:hybridExecutor} shows some general details of its implementation. As it can be seen, we implemented a naive policy for the hybrid executor. It will schedule the task depending on the current local load: if there are pending tasks in the local thread pool queue (line \ref{line:hybridExecutor:idle}), new tasks are scheduled to be run as serverless functions (line \ref{line:hybridExecutor:serverless}). But whenever the current local load is capable of executing another task, it will be executed locally (line \ref{line:hybridExecutor:local}).

The idea behind the \texttt{HybridExecutor} is to have a VM capable of managing a baseline number of simultaneous tasks, but, at the same time, be able to scale it vertically with cloud functions automatically and without the need for prior provisioning. This allows us to have, at the same time, the benefits of a monolithic system together with the best of the serverless paradigm: the cost of the VM is constant and low, but we have the elasticity and immediate scalability of serverless. 

\begin{lstlisting}[language=Java, caption=HybridExecutor implementation, label={lst:hybridExecutor}, escapechar=@]
public abstract class ServerlessHybridExecutorService implements ExecutorService {
    
    private ExecutorService localExecutorService;
    private ExecutorService serverlessExecutorService;
    ...

    public <T> Future<T> submit(Callable<T> task) {
    if (task == null) 
        throw new NullPointerException();

    if (!(task instanceof Serializable))
        throw new IllegalArgumentException("Tasks must be Serializable");

    Future<T> f = null;
    if (isLocalExecutorIdle()){ @\label{line:hybridExecutor:idle}@
        f = localExecutorService.submit(task); @\label{line:hybridExecutor:local}@
    } else {
        Callable<T> localCallable = () -> {
            ThreadCall call = new ThreadCall("ServerlessExecutor-" 
                                + Thread.currentThread().getName());
            call.setTarget(task);
            return invoke(call);
        };
        f = serverlessExecutorService.submit(localCallable); @\label{line:hybridExecutor:serverless}@
    }
    submittedTasks.add(f);
    return f;
    }
    ... 
}
\end{lstlisting}

\subsection{Limitations of our approach} 
As of today, our serverless executor model presents some limitations. As every serverless approach, it has to deal with various constraints of serverless computing, namely limited duration of functions, limit on concurrency and invocation frequency, lack of direct network connectivity between functions, and the stateless nature of functions.

\paragraph{Limitation \#1} The evolution of the industry in cloud services allows us to be optimistic when it comes to the duration of the functions, limit of concurrency and frequency of invocation. We can observe that these limits are getting higher and, over the time, will have less and less relevance. In our case, they have been taken into account and an adaptation has been made to overcome some of these. For example, we need to control the concurrency of cloud functions to avoid exceeding concurrency limits. We also have to tune some parameters of the implemented algorithms to ensure a proper duration of tasks: too short tasks may not be able to mask cloud functions overheads. Some of these adaptations could be avoided using more sophisticated scheduler policies for both the serverless executor and the hybrid executor, but that is out the scope of this article and will remain as an open challenge for future work.

\paragraph{Limitation \#2: Stateful computations} As of today, our approach is stateless. That is, the tasks launched as cloud functions do not share data with one another. Instead, all data will be received and returned as parameters. This limits the range of algorithms that can be analyzed. However, it is an intentional decision to start with a simpler and pure-serverless approach. We leave for future work the adoption of a stateful serverless architecture that will be needed to support other algorithms. Note that Crucial \cite{crucial} already offers the possibility to have a shared state using a distributed shared object (DSO) layer so transition to other kind of algorithms should be relatively simple. 

\section{Methodology}
\label{sec:methodology}

\subsection{Algorithms}

In this work we target three of these irregular and unbalanced algorithms: UTS, Mariani-Silver and Betweenness Centrality (BC).

\subsubsection{UTS}

The Unbalanced Tree Search (UTS) benchmark~\cite{uts} was initially proposed by Prins et al.~\cite{prins03} in 2003 and has since been used as an irregular application for evaluating load-balancing algorithms in several parallel computing architectures and using different programming models like MPI~\cite{DinanOSPST07}, Unified Parallel C~\cite{OlivierP08} or X10's APGAS~\cite{lifelineBasedGLB}. However, all these programming models are not elastic. The implementation of UTS presented in this work is the first that tackles an elastic environment.

UTS counts the number of nodes in a highly unbalanced task tree that is dynamically generated using SHA-1 cryptographic hashes. The number of children of a node is a random variable with a given distribution. In this paper we used a geometric distribution with an expected branching factor $b_{0}=4$ and a depth cut-off $d$ between $14$ and $18$. Table~\ref{tab:treeSize} shows how the number of nodes increases exponentially as we increase the depth cut-off. 

\begin{table}[!t]
  \caption{
    UTS tree size for seed $=19$ and $b_{0}=4$
  }
  \label{tab:treeSize}
  \centering
  \begin{tabular}{cr}
    Depth $d$ &  Tree size (\# of nodes) \\
    \hline
    $14$ &   $1,057,675,516$  \\
    $15$ &   $4,230,646,601$  \\
		$16$ &  $16,922,208,327$  \\
		$17$ &  $67,688,164,184$  \\
		$18$ & $270,751,679,750$  \\
    
    \bottomrule
  \end{tabular}

\end{table}

Each parallel task iterates through a maximum of $n$ nodes of the assigned subtree. Due to the imbalance of UTS, some tasks will traverse $n$ nodes while many others will traverse significantly less. Another parameter that can be configured is the split factor: the maximum amount of partitions a subtree is divided into before launching new parallel tasks.
In listing~\ref{lst:uts} we see that each remote task (line~\ref{line:uts:remote}) receives a \texttt{bag} parameter that encapsulates the subtree assigned to the task. Once the \texttt{bag} is processed, it is returned (line~\ref{line:uts:returnBag}) to the master program where it is added to a queue (line~\ref{line:uts:addQueue}). Another thread is responsible to retrieve bags from the queue (line~\ref{line:uts:pollQueue}), resizing subtress (line~\ref{line:uts:resize}), and subsequently launch new parallel tasks (line~\ref{line:uts:parallelize}).

\begin{lstlisting}[language=Java, caption=UTS serverless implementation, label={lst:uts}, escapechar=@]
private void uts(List<Bag> bags) { 
	parallelize(bags, iters);

	while(<still active threads || queue not empty>) {
		Bag bag = queue.poll();  @\label{line:uts:pollQueue}@
		if (bag != null) {
			activeThreads.addAndGet(-1);
			bags = resizeBag(bag, splitFactor); @\label{line:uts:resize}@
			parallelize(bags, iters); @\label{line:uts:parallelize}@
		}
	}
}

private void parallelize(List<Bag> bags, int iters) {
	activeThreads.addAndGet(bgas.size());
	
	for (Bag bag : bags) {
		Future<Object> future = localExecutor
			.submit(new LocalUTSCallable(bag, iters));
	}
}

class LocalUTSCallable implements Callable<Object> {
	...
	public Object call() throws Exception {
		Future<Result> future = serverlessExecutor.submit(
			new RemoteUTSCallable(bag, iters));
		Result result = future.get();
		queue.offer(result.getBag()); @\label{line:uts:addQueue}@
	}
}

class RemoteUTSCallable implements Callable<TMResult>, Serializable { @\label{line:uts:remote}@
	...
	public Result call() throws Exception {
		// initialization of MessageDigest
		...
		for (int n = numberOfIterations; n > 0 && bag.size > 0; --n) {
			// traverse one node
			...
		}
		return new Result(bag); @\label{line:uts:returnBag}@
	}
}
\end{lstlisting}

\subsubsection{Mariani-Silver}

The simplest algorithm to render the Mandelbrot Set is the naive Escape Time algorithm, where a repeating calculation is performed for every point in the plot area. However, there are several optimization techniques that can be applied to speed up the rendering without having to compute all the points.
One of these techniques is the Mariani-Silver algorithm, an adjacency optimization technique. This algorithm relies on the fact that the Mandelbrot set is connected: there is a path between any two points belonging to the set. In this algorithm, a rectangular grid is recursively subdivided into subrectangles. The pixels on the boundary of each subrectangle are evaluated, and if they all evaluate to the same result the subrectangle is filled with the same solid color; otherwise the subrectangle is divided again into two or more smaller rectangles. The algorithm is recursively applied to each piece until the maximum nesting depth is reached. In this case, all pixels of the rectangle are evaluated. 

Listing~\ref{lst:mariani} shows the main code of the Mariani-Silver algorithm. When the rectangle has to be split (line~\ref{line:mariani:split}), new recursive tasks are generated. This kind of nested loops are commonly found in algorithms using hierarchical data structures, such as adaptive meshes, graphs and trees, and also in algorithms like this that uses recursion, and that parallelism can be exploited at each level of recursion. It is difficult to avoid the nested loops, and therefore these algorithms exhibit irregular workloads.

\begin{lstlisting}[language=Java, caption=Mariani-Silver Callable code, label={lst:mariani}, escapechar=@]
public Result call() {
   Result result = new Result(rectangle);
   if (borderHasCommonDwell(rectangle)) {
      result.setNextAction(Result.Action.FILL);
      result.setDwellToFill(rectangle.getBorderDwell());
   } else if (rectangle.getDepth() >= MAX_DEPTH) {
      // per-pixel evaluation of the rectangle
      int[][] dwellArray = evaluate(rectangle);
      result.setNextAction(Result.Action.SET_DWELL_ARRAY);
      result.setDwellArray(dwellArray);
   } else {
      result.setNextAction(Result.Action.SPLIT); @\label{line:mariani:split}@
      // New tasks will be generated
   }
   return result;
}
\end{lstlisting}

Each parallel task evaluates a single subrectangle and returns the action to take after the evaluation. If the subrectangle has to be subsequently split, the master will be responsible of creating and invoking the new tasks. Otherwise, the evaluation returns the dwell values to assign to the subrectangle (either the boundary value or all individual values of the subrectangle). The maximum dwell, the number of initial subdivisions, the split factor, and the maximum depth is configurable.

\subsubsection{Betweenness Centrality}
\label{sec:bc}

Betweenness Centrality (BC) is a graph algorithm that computes a centrality metric for each vertex of a graph.
In particular, the version used in this work is taken from SSCA2 (Scalable Synthetic Compact Application 2) v2.2 benchmark~\cite{ssca2}. Our Java implementation is based on the X10 code used to validate the GLB programming model for large-scale distributed systems~\cite{glbPPAA14}.

Let $\sigma_{st}$ denote the number of shortest paths between vertices $s$ and $t$, and $\sigma_{st}(v)$ the number of those paths passing through $v$. Betweenness Centrality of a vertex $v$ is defined as
\begin{equation}
  BC(v) = \sum_{s \ne v \ne t \in V} \frac{\sigma_{st}(v)}{\sigma_{st}}.  
  \label{eq:BCvertex}
\end{equation}
The output of the algorithm is a betweenness centrality score for each vertex in the graph.

The implementation follows the Brandes' algorithm described in the benchmark, augmenting Dijkstra's single-source shortest paths (SSSP) algorithm, for unweighted graphs. 
Listing \ref{lst:bc} shows our serverless implementation, which assumes that the whole graph fits in the memory of each function, allowing the same graph to be generated across all functions (line \ref{line:bc:functionGraph}). The set of $N$ vertices is statically partitioned among $T$ tasks (line \ref{line:bc:LocalCallable}), so each function is responsible for performing the computation for the source vertices assigned to one of the tasks (for all N target vertices) and computes its task-local \texttt{betweennessMap} (line \ref{line:bc:computeBC}). Then, all local \texttt{betweennessMaps} are collected as results of the serverless functions and are finally aggregated in a \texttt{globalBetweennessMap} (line \ref{line:bc:globalmap}).

As the graph is generated (line \ref{line:bc:generateGraph}) following a “recursive matrix” (\mbox{R-MAT}) model~\cite{rmat}, the amount of work to compute each source vertex is different. The vertices are permutated (line \ref{line:bc:permuteVertices}) before partitioning the graph in order to create more homogenous tasks, but the resulting tasks are still not perfectly balanced.

\begin{lstlisting}[language=Java, caption=Betweenness Centrality serverless implementation, label={lst:bc}, escapechar=@]
  private void BetweennessCentrality(){
    Rmat rmat = new RMat(...);
    setup(rmat);
    runServerless();
  }

  private void setup(RMat rmat) {
    graph = rmat.generate();  @\label{line:bc:generateGraph}@
    graph.compress();
    N = graph.numVertices();
    permuteVertices(); @\label{line:bc:permuteVertices}@
    betweennessMap = new double[N];
  }

  private void runServerless() {
    List<Future<Object>> futures = new ArrayList<>();
    for (int startIndex = 0; i < N; startIndex += taskSize) {        
      Future<Object> f = localExecutor.submit(
          new LocalCallable(startIndex, startIndex + taskSize - 1, rmat));  @\label{line:bc:LocalCallable}@
      futures.add(f);
    }
  }

  class LocalCallable implements Callable<Object> {
    ...
    public Object call() throws Exception {
      Future<Result> f = serverlessExecutor.submit(
          new ServerlessCallable(startIndex, endIndex));
      Result result = f.get();      
      
      double[] bMapLocal = result.getBetweennessMap();
      for(int i = 0; i < N; i++) {
        if (bMapLocal[i] != 0) {
          betweennessMap[i] = bMapLocal[i]; @\label{line:bc:globalmap}@
        }
      }
      return null;
    }
  }  

  public class ServerlessCallable implements Callable<Result>, Serializable  {
    ...
    public Result call() throws Exception {
        generateGraph(); @\label{line:bc:functionGraph}@
        setArrays();
        for(int i = startIndex; i <= endIndex; i++){
            computeBC(i); @\label{line:bc:computeBC}@
        }
        Result result = new Result(betweennessMapLocal);
        return result;
    }
  }
\end{lstlisting}

\subsection{Algorithms characterization}

All the studied algorithms have subtasks with irregular computation load but each one has specific characteristics. Table~\ref{tab:characterization} summarizes some of these differences. For example, UTS and Mariani-Silver are recursive algorithms that generate data dependencies between tasks, whereas BC workload can be partitioned in independent tasks beforehand.

\begin{table*}[t]
\caption{Characterization of the tested algorithms.}
\label{tab:characterization}
\begin{center}
    \begin{tabular}{p{2.2cm} p{2.2cm} p{2.6cm} p{3.5cm} p{3.0cm} p{3.0cm}}
    \hline
    Algorithm & Task\newline dependencies & Coefficient of\newline Variation ($C_L$) & Parameters & Input & Output \\ \hline
    
		\textit{UTS} & Yes & $1.20$ & Seed $= 19$\newline$b_0 = 4$\newline$d=18$ & Tree parameters & Node counter \\ \hline
		\textit{Mariani-Silver} & Yes & $4.06$ & Width $=4096$\newline Height$=4096$ & Rectangle parameters & Color matrix \\ \hline 
		\textit{Betweenness Centrality} & No & $0.23$ & Seed $=2$ \newline $T=128$\newline R-MAT $= \left(\begin{matrix}0.55 & 0.1 \\ 0.1 & 0.25\end{matrix}\right)$& Unweighted Graph & Centrality array \\ \hline
		
    \end{tabular}
\end{center}

\end{table*}

An imbalance metric is important to characterize irregular algorithms. A typical metric is the standard deviation of the execution times of all subtasks. The reasoning behind this metric is that if
workloads are evenly distributed, the standard deviation will be small. The smaller this metric, the greater the
balance of the algorithm tasks. Considering $L$ as the set of execution times of all subtasks of an algorithm, an imbalance metric can then be defined in terms of the coefficient of variation of $L$:

\begin{equation}
C_L = \frac{\sigma_L}{\mu_L}
\label{eq:coefficientVariation}
\end{equation}

As shown, $C_L$ is defined as the ratio of the standard deviation $\sigma_L$ over the mean $\mu_L$. $C_L$ expresses the extent of variability in relation to the mean of the measured values without depending on the
measurement unit. This metric has been used to quantify load imbalance on virtualized enterprise servers~\cite{arzuaga} as well as in many other areas of computer science. The higher the $C_L$, the greater the variation in measured values, and thus greater the imbalance of the algorithm. Table \ref{tab:characterization} also contains the coefficient of variation $C_L$ for each one of the studied algorithms. Mariani-Silver has the highest $C_L$ coefficient, which implies that it has the highest task duration variability, and the BC algorithm has the lowest $C_L$.

\begin{figure*}
  \centering
  \begin{subfigure}[b]{0.3\textwidth}
      \centering
      \includegraphics[width=\textwidth]{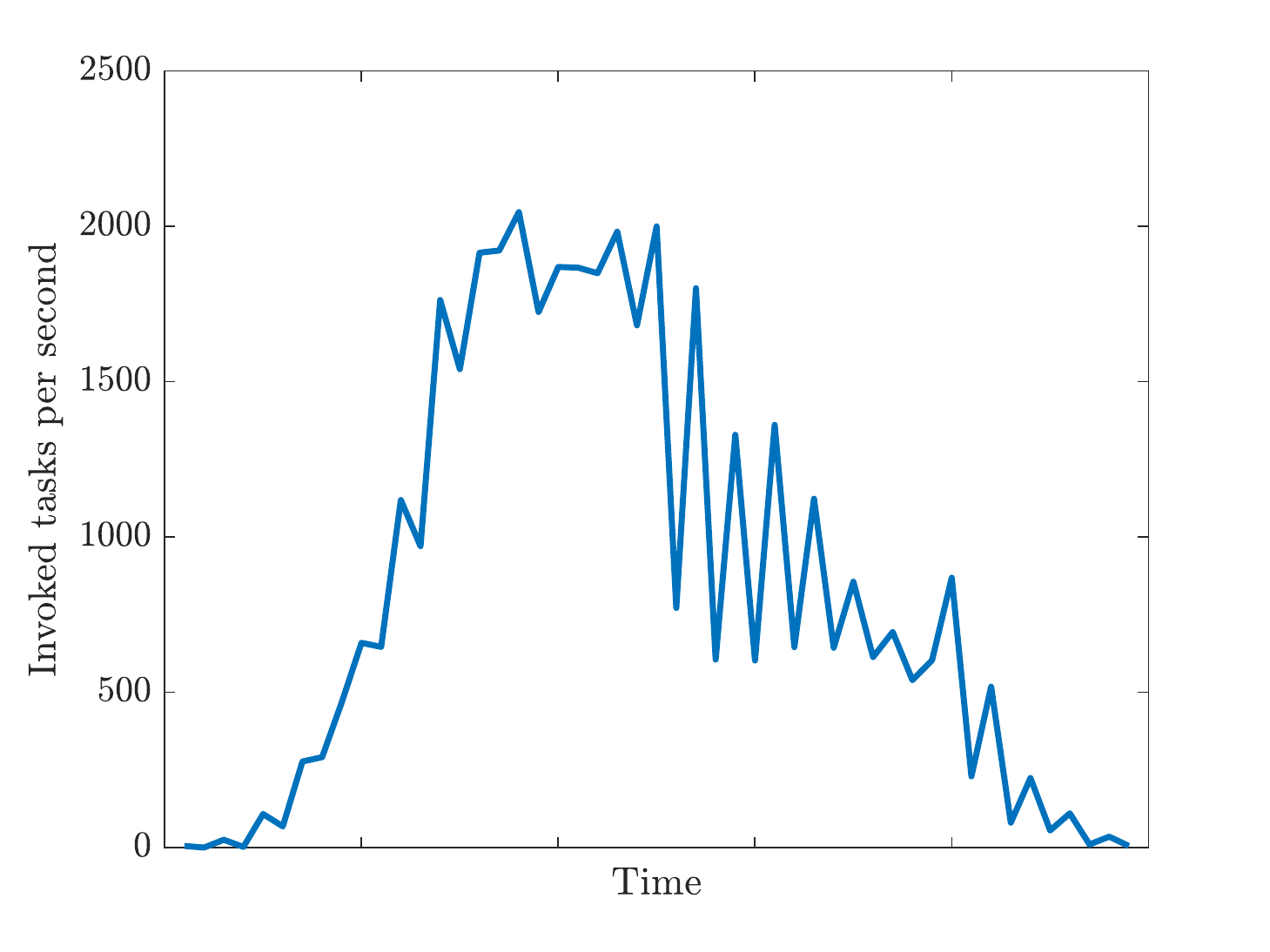}
      \caption{UTS}
      \label{fig:UTSRate}
  \end{subfigure}
  \hfill
  \begin{subfigure}[b]{0.3\textwidth}
      \centering
      \includegraphics[width=\textwidth]{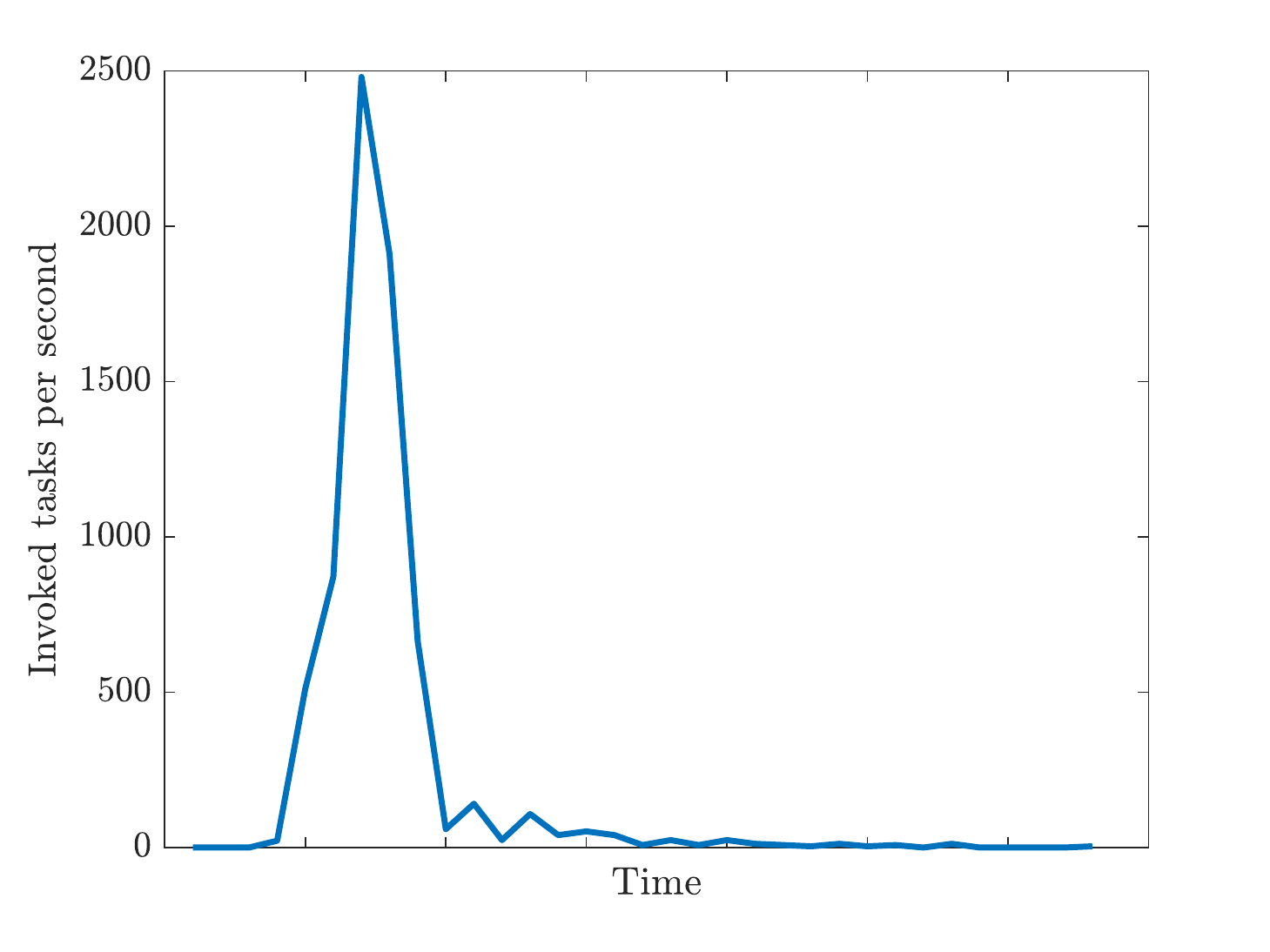}
      \caption{Mariani-Silver}
      \label{fig:MarianiSilverRate}
  \end{subfigure}
  \hfill
  \begin{subfigure}[b]{0.3\textwidth}
      \centering
      \includegraphics[width=\textwidth]{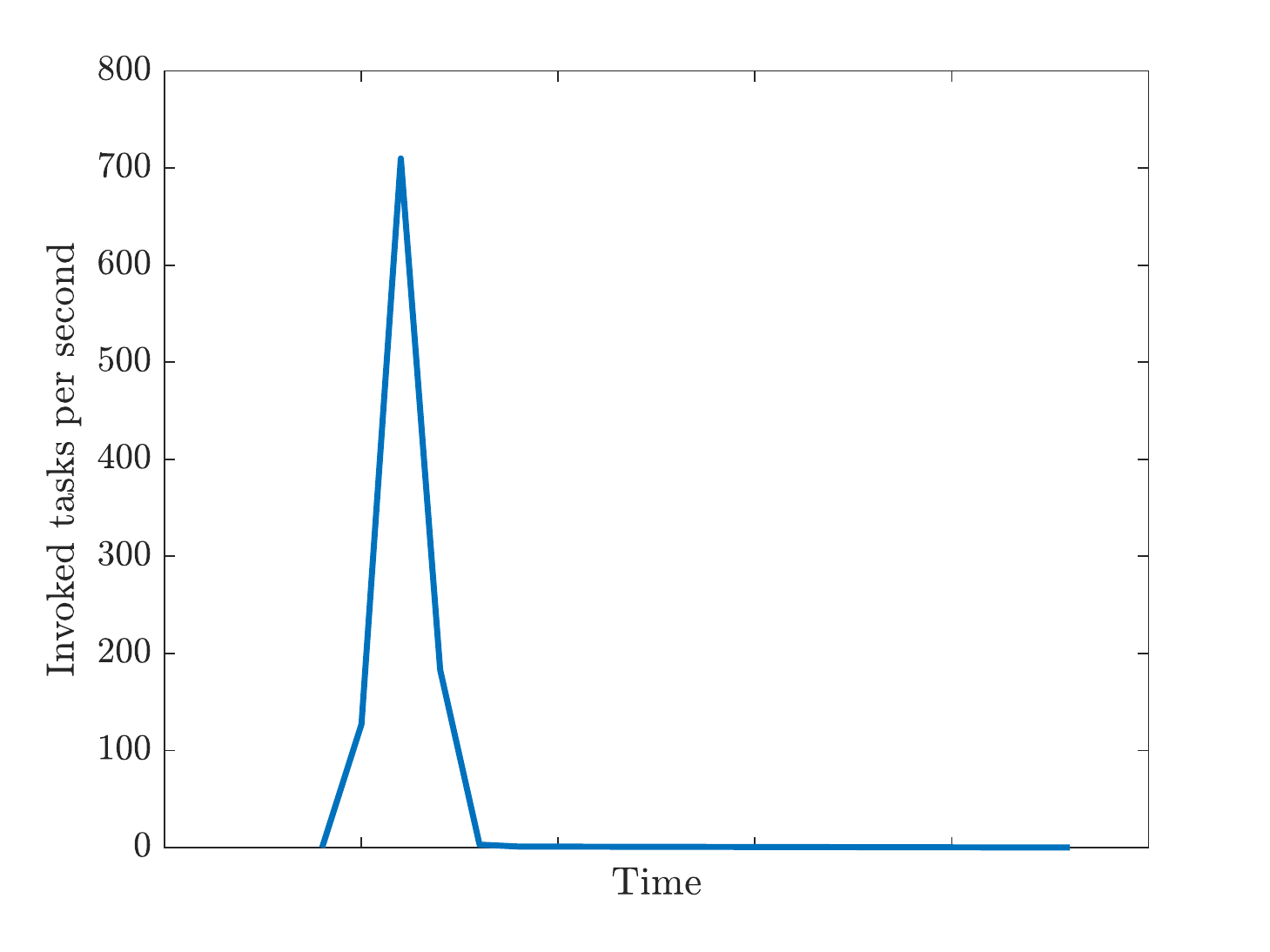}
      \caption{Betweenness Centrality}
      \label{fig:BCRate}
  \end{subfigure}
  \caption{Task generation rate}
  \label{fig:taskGenerationRate}
\end{figure*}

The management of resources is, in part, influenced by the number of task generated during the execution of the algorithms. Therefore, another important factor to take into account is the task generation rate. Figure \ref{fig:taskGenerationRate} shows how many tasks per second are generated by each algorithm. We can see how UTS has a very irregular high task generation rate during practically the entire execution. This is the opposite of the case of the BC, where all the tasks are generated at once at the beginning of the execution. Mariani-Silver is an intermediate term, it generates a large part of the tasks at the beginning, but during the rest of the execution it maintains an erratic rhythm on a much smaller scale.

\begin{figure*}
  \centering
  \begin{subfigure}[b]{0.3\textwidth}
      \centering
      \includegraphics[width=\textwidth]{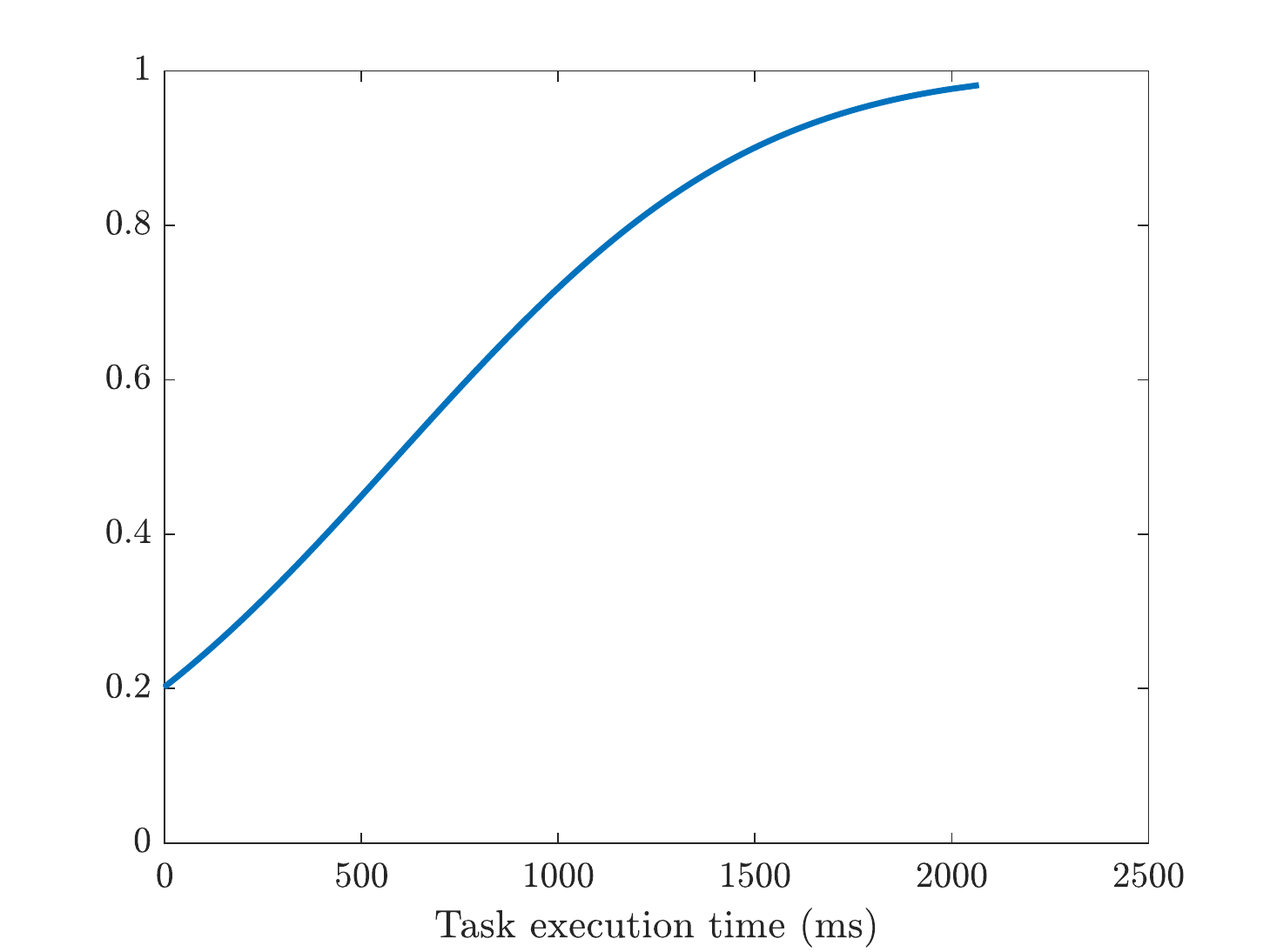}
      \caption{UTS}
      \label{fig:cdfUTS}
  \end{subfigure}
  \hfill
  \begin{subfigure}[b]{0.3\textwidth}
      \centering
      \includegraphics[width=\textwidth]{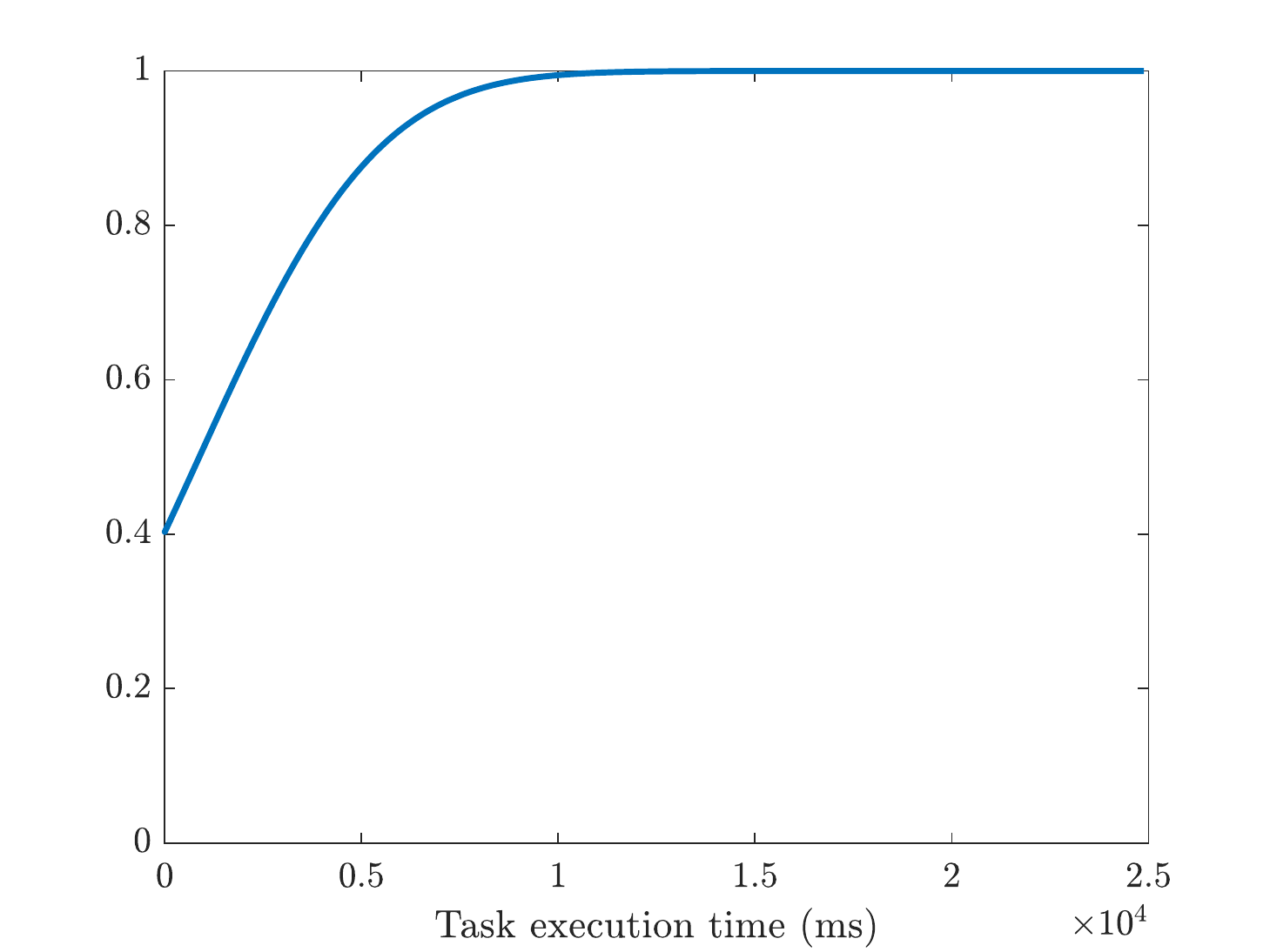}
      \caption{Mariani-Silver}
      \label{fig:cdfMariani}
  \end{subfigure}
  \hfill
  \begin{subfigure}[b]{0.3\textwidth}
      \centering
      \includegraphics[width=\textwidth]{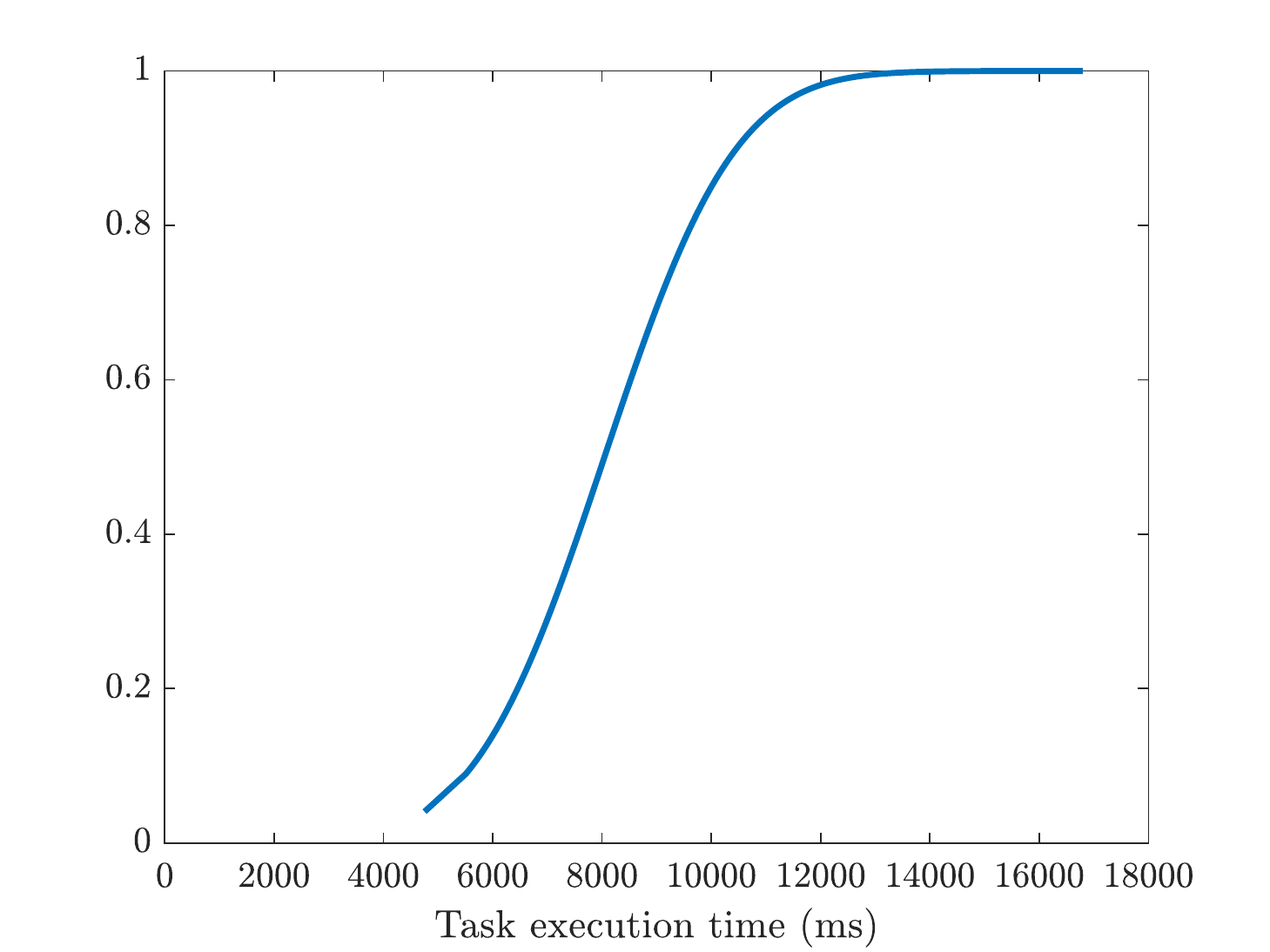}
      \caption{Betweenness Centrality}
      \label{fig:cdfBC}
  \end{subfigure}
  \caption{Cumulative Distributive Function (CDF) of the execution time of each task}
  \label{fig:CDFexecution-time}
\end{figure*}

In order to complement the characterization provided by the $C_L$ coefficient, the Cumulative Distributive Function (CDF) of the execution time of the tasks of each algorithm has been studied. Figure \ref{fig:CDFexecution-time} contains the graphical representation of the CDF for each algorithm. We can see that, in UTS, 20\% of the tasks have a duration of less than 1ms and from the duration of the tasks it is uniformly distributed until reaching 2000ms. In the case of Mariani-Silver we can observe a very wide task duration interval, where 40\% of the tasks are under 1s and 99\% are under 10s. The large percentage of tasks with a very short duration is due to the Mariani-Silver adjacency optimization technique. This technique allows to safely fill in the entire region without computing the interior when all the boundary pixels of the region are  evaluated to the same result. In the Mariani-Silver's CDF, the long tail of the graph on the right side is particularly striking, indicating the existence of some (very few) tasks with a very high execution time (up to 25s) compared to the rest of the tasks. The BC's CDF shows a more homogenous task duration than the other two algorithms, matching the $C_L$ coefficient. Even so, execution times are evenly distributed between 4 s and 12 s for most tasks, which is still a wide range despite the use of random shuffling of the nodes assigned to each task.

\subsection{Performance and cost metrics}

When considering performance of a serverless environment, a basic performance indicator is the total execution time. Also, as the selected algorithms can be quantitatively characterized by the number of traversed nodes (UTS) or the number of computed points (Mandelbrot), we can define a throughput indicator dividing the number of nodes or points by the total execution time.

Regarding cost, we calculate the cost metric using the serverless provider listed prices. The total cost of running an algorithm with serverless functions is calculated as: the cost of the invocations, plus the cost of the execution time, plus the cost of the VM where the client application is executed (Eq. \ref{eq:costServerless}):

\begin{equation}
\text{Cost}_{\text{Serverless}}=\text{Cost}_{\text{Invocations}} + \text{Cost}_{\text{Execution}} + \text{Cost}_{\text{Client}}
\label{eq:costServerless}
\end{equation}

Considering current AWS prices, the total cost of $n$ invocations is calculated as:

\begin{equation}
\text{Cost}_{\text{Invocations}}= \lambda_i \cdot n
\label{eq:costInvocations}
\end{equation}

Where $\lambda_i$ is the AWS Lambda invocation cost per serverless function. The cost of the execution time is the gigabyte-second price multiplied by the gigabytes of memory assigned to functions and the billed duration of each execution in seconds ($t_i$): 

\begin{equation}
\text{Cost}_{\text{Execution}}= \lambda_e \cdot \frac{\text{memory in MB}}{1024 \text{ MB}} \cdot \sum_{i=0}^{n}{t_i}
\label{eq:costExecution}
\end{equation}

Where $\lambda_e$ is the AWS Lambda execution cost per Gigabyte per second. The cost of the client VM can be computed as the VM price per hour divided by 3600 (seconds per hour) and multiplied by the total execution time (in seconds):

\begin{equation}
  \text{Cost}_{\text{Client}} = \frac{\text{VM Price }\$}{3600} \cdot t_{\text{total}}
\end{equation}

Table \ref{tab:awscost} shows the specific AWS prices from previous formulas at the time of performing the experiments.

\begin{table}
  \caption{AWS Lambda prices.}
  \centering
  \begin{tabular}{c | r}
    Constant & Value \\ \hline
    $\lambda_i$ & \$$0.0000002$ \\
    $\lambda_e$ & \$$0.0000166667$ \\
    VM Price (m5.xlarge) & \$$0.192$ \\
    VM Price (c5.2xlarge) & \$$0.34$
  \end{tabular}
  \label{tab:awscost}
\end{table}

To discuss trade-offs between cost and performance, we also refer to the price to performance ratio ($R_{\text{Price-Performance}}$), a useful metric to show the throughput that can be achieved per dollar spent:

\begin{equation}
R_{\text{Price-Performance}} = \frac{\text{Throughput}}{\text{Cost}_{\text{Execution}}}
\label{eq:ratioPricePerformance}
\end{equation}

\subsection{Experimental setup}

All experiments in this paper are conducted in Amazon Web Services (AWS).
We use AWS Lambda as serverless functions, with each function configured with the amount of memory to have the equivalent to one full vCPU of compute time, according to AWS documentation \cite{lambdaMemory}. If not stated otherwise, the client application is executed in a \texttt{m5.xlarge} instance in the same region assigned to AWS Lambda. All execution times are the mean of at least 10 runs.

\section{Performance analysis}
\label{sec:performance}

In terms of performance, current serverless computing platforms have some limitations that must be considered beforehand. One of the main problems of serverless applications is the predictability of performance~\cite{berkeleyServerless}. One source of unpredictability is the known problem of cold starts of cloud functions: the time it takes to start a new function container, to initialize the software environment, and to initialize the specific user code. 
Depending on the application, this startup latency may be high. 
In this sense, the advantage of task-parallel applications is that these cold starts only affect the first batch of functions, and successive tasks benefit from having already warm containers.

Obviously, the predictability of performance in FaaS is lower than in traditional IaaS VM instances, and even lower than in HPC dedicated clusters in computing centers. However, there is an increasing interest in moving HPC workloads to the cloud, and particularly to FaaS platforms, due to its elasticity, availability, and pay-as-you-go model.\cite{FuncX}

For task-parallel workloads in FaaS, one of the main factors that affect final performance is the overhead of running a remote task. 
Table~\ref{tab:overheads} compares the overheads to invoke a dummy task as a serverless function and as a local thread, using the executors presented in this paper. 
The task simply receives an input parameter and returns an output parameter without performing intensive computations. 
Both parameters are short strings. 
To obtain an average overhead, we measure the time to sequentially submit and obtain results of 1k serverless functions and 1M local threads. 
Both executors are previously warmed up to discard overheads attributed to thread pool initialization, connection set up and cold starts. 
The experiments are carried out in an EC2 \texttt{c5.2xlarge} instance, and we use AWS Lambda to run serverless functions.
Even though the overhead of a serverless function ($\sim13$ms) is three orders of magnitude greater than the overhead of a local thread, it is still low enough to permit efficient execution of parallel workloads.

\begin{table}[!t]
  \caption{
    Average invocation overheads
  }
  \label{tab:overheads}
  \centering
  \begin{tabular}{lc}
    &  Overhead   \\
    \hline
    Serverless Functions & \multicolumn{1}{r}{$13$ms}  \\
    Local threads & \multicolumn{1}{r}{$18 \mu$s}  \\
    
    \bottomrule
  \end{tabular}

\end{table}

\begin{table*}[!t]
  \caption{
    Performance and parallel efficiency of UTS.
  }
  \label{tab:parallelEfficiency}
	\centering
	\begin{tabular}{p{4.5cm}rrrrr}
    & Logic CPUs  &  Depth & Time (s) & Throughput (M nodes/s) & Parallel efficiency  \\
    \midrule
		Sequential & $1$ & $14$ & $75.86$ & $13.94$ &  \\ \hline
		AWS Lambda & $96$ & $17$ & $74.55$ & $907.94$ & $67.84\%$ \\ 
		AWS Lambda & $48$ & $17$ & $141.67$ & $477,76$ & $71.40\%$ \\ 
    \hline
    \texttt{c5.24xlarge} & $96$ & $17$ & $113.90$ & $594.22$ & $44.40\%$ \\ 
		\texttt{c5.24xlarge} (HT disabled) & $48$ & $17$ & $131.73$ & $513.82$ & $76.78\%$ \\ 
		\texttt{c5.12xlarge} & $48$ & $17$ & $214.83$ & $315.06$ & $47.09\%$ \\ 
    \bottomrule
  \end{tabular}
\end{table*}

\subsection {Effects of Hyper-Threading to compute-intensive tasks}
\label{sec:hyperthreading}

Running a parallel algorithm in a distributed environment, either serverless functions or a cluster, obviously adds some overheads to the final execution time, mainly related to serialization and network latency. If permitted by the scale of the problem, a viable alternative is to run a parallel version of the code in one of the largest virtual machine instances offered by cloud providers. In Table~\ref{tab:parallelEfficiency} we show a comparison of the performance achieved with the serverless and parallel versions of UTS. The parallel version is run in a \texttt{c5.24xlarge} VM with $96$ vCPUs and in a \texttt{c5.12xlarge} VM with $48$ vCPUs. For comparison purposes, the serverless version has been limited here to launch a maximum of $96$ or $48$ concurrent functions. In order to have a reference of the throughput achieved, we also run a single-threaded version in the same VM using a lower depth parameter of $14$. The selected depth does not affect the throughput of the sequential code.

Despite the inherent overheads, the serverless version appears to be surprisingly faster than the parallel version: traversing a 17-depth tree takes $74.55$s in average with a maximum of $96$ concurrent AWS Lambda functions while it takes $1.5x$ more time with the \texttt{c5.24xlarge} VM.
Comparing the throughput of the parallel version with the sequential version, we see that the parallel efficiency is very low ($44.4\%$).
This can be attributed to the effect of Intel's Hyper-threading (HT) to a compute-intensive task like this. 
Hyper-Threading technology makes a single physical core appear as two logical cores. The physical core resources are shared and the architectural state is duplicated for the two logical cores~\cite{marr2002hyper}. It is known that depending on the characteristics of the application run, Hyper-Threading may help or hinder performance~\cite{tau2002empirical}. In particular, compute-intensive applications have less chance to be improved in performance from Hyper-Threading because the CPU resources could already be highly utilized~\cite{tau2002empirical}.
When running the parallel version in the VM, each pair of vCPUs or logical CPUs share the same physical core. In effect, we verified that disabling Hyper-Threading, and thus reducing in half the number of logical CPUs, the throughput of the parallel version only decreases slightly and the parallel efficiency reaches a more acceptable rate of $76.78\%$. 

To finish the study of the parallel efficiency we can make the comparison in the case of a parallelism of 48. In this case we have a VM with HyperThreading (c5.12xlarge), a VM without HyperThreading (c5.24xlarge) and the AWS Lambda limited to the maximum use of 48 concurrent functions. In this scenario, the comparison between the two VMs is not interesting, since c5.24xlarge has twice as many physical cores as c5.12xlarge. However, both have been configured to have the same number of logical cores, just like the serverless based system. This allows us to contextualize the parallel efficiencies of the serverless functions and measure the effects of the existing overheads.

Results show that the \texttt{c5.24xlarge} VM with HyperThreading disabled gets the best performance, with a execution time of $131.73$s. However, the execution time of the serverless functions is only $7\%$ higher ($141.67$s), in contrast with the c5.12xlarge VM execution time, which is $51\%$ higher ($214.83$s) than the serverless functions . Therefore, the functions get a parallel efficiency much closer to the parallel efficiency of a VM without HyperThreading.

This is an interesting insight that must be considered when comparing performance in virtual machines versus serverless functions. Although, according to AWS documentation, each Lambda function has been configured to an equivalent of one vCPU~\cite{lambdaMemory}, a direct comparison to a VM with the same theoretical resources may lead to misleading results, especially for compute-intensive tasks. While all physical cores of the VM are at full utilization, the serverless functions are run in AWS Lambda infrastructure, a setting that is out of the control of the user and its core real utilization can vary and is not known by the user either. However, this experiment suggests that, at least at AWS Lambda, compute intensive parallel tasks may exhibit better performance than the same tasks at a VM with the same resources in terms of vCPUs.

\subsection{UTS Optimizations}

Our serverless implementation of UTS can be optimized to further reduce total execution time. If the application can sustainably maintain a high task concurrency, UTS tree traversal finishes earlier. The concurrency greatly depends on the split factor, the number of parts that a task is split into. A high split factor generates more tasks, but an excessive number of tasks is counterproductive because adds overheads. Another parameter that can be tuned is the number of nodes processed at each task.

The basic serverless implementation maintains these two parameters static during the execution. Here, we apply an optimization to dynamically modify the split factor and the number of nodes processed depending on the current level of concurrency. 

\begin{lstlisting}[language=Java, caption=UTS Optimization example, label={lst:uts_optimization}, escapechar=@]
  private void uts(List<Bag> bags) { 
    parallelize(bags, iters);
    iters = 50_000; @\label{line:uts_optimization:low_iters}@
    splitFactor = 200; @\label{line:uts_optimization:high_split_factor}@
  
    while(<still active threads || queue not empty>) {
      Bag bag = queue.poll();  
      if (bag != null) {
        activeThreads.addAndGet(-1);

        if (step==0 && active threads > 800) {
          step++;
          splitFactor = 50; @\label{line:uts_optimization:med_split_factor}@
          iters = 2_500_000;
        }
        if (step==1 && active threads > 1300) {
          step++;
          splitFactor = 5; @\label{line:uts_optimization:low_split_factor}@
          iters = 5_000_000; @\label{line:uts_optimization:high_iters}@
        }
        if (step==2 && active threads < 1100) {
          step++;
          iters = 2_500_000;
        }
        if (step==3 && active threads < 100) {
          step++;
          iters = 1_000_000; @\label{line:uts_optimization:low_iters_final}@
        }

        bags = resizeBag(bag, splitFactor); 
        parallelize(bags, iters);
      }
    }
  }  
\end{lstlisting}

In order to achieve a significant improvement the parameters updates do not need to be done with any complex parameter curve. Listing \ref{lst:uts_optimization} shows an example for the UTS optimization implementation, where simple parameters updates are performed in four different stages of the UTS execution. We assign a large split factor (line \ref{line:uts_optimization:high_split_factor}) until a high concurrency is achieved, and then slowly start to decrease the split factor (lines \ref{line:uts_optimization:med_split_factor} and \ref{line:uts_optimization:low_split_factor}) as the measured concurrency also goes down defined levels. Likewise, the number of processed nodes is kept lower at the beginning (line \ref{line:uts_optimization:low_iters}) and the end of the execution (line \ref{line:uts_optimization:low_iters_final}), and is increased when high concurrency levels are achieved (line \ref{line:uts_optimization:high_iters}). 

\begin{figure}[h]
  \centering
  \includegraphics[width=0.9\linewidth]{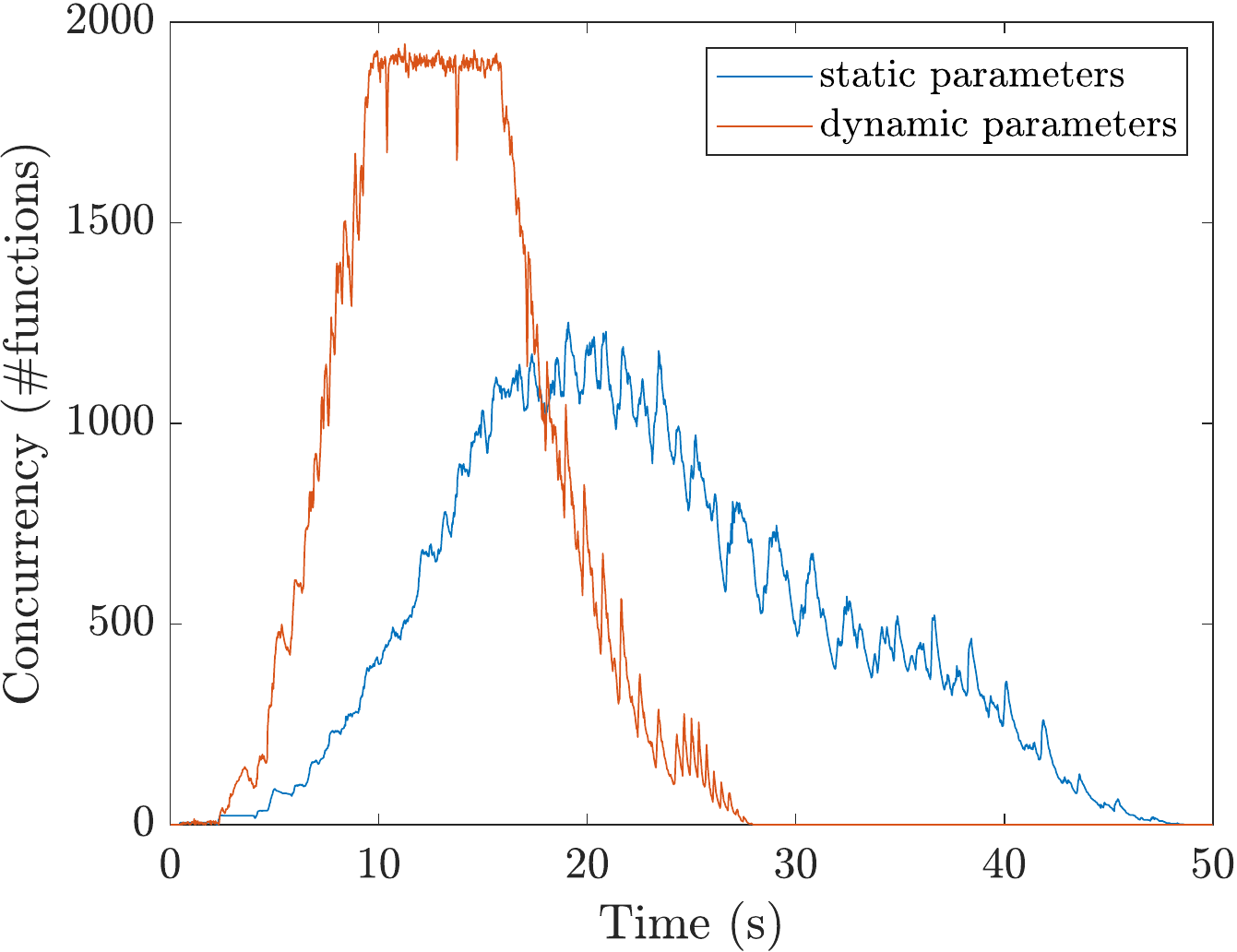}
  \caption{
    Concurrency and total execution time for serverless version of UTS ($d=18$).
  }
  \label{fig:uts18concurrencyOptimized}
\end{figure}

Figure~\ref{fig:uts18concurrencyOptimized} shows the concurrency and total execution time improvement after this optimization is applied. The maximum concurrency allowed by AWS Lambda in this environment is $2,000$ functions, so we configure the serverless thread pool with this maximum. In the dynamic version, we observe that the effective concurrency saturates near this maximum, achieving higher concurrency than the static version. Thanks to this, this version finishes in $27.9$s.

\subsection{Preliminary hybrid executor evaluation}

To assess the performance implications of the hybrid executor, we conduct some preliminary evaluations using the Mariani-Silver algorithm 
to generate a $4096\times4096$ pixels image of the Mandelbrot Set, where the fractal space is divided adaptively using dynamic parallelism. To maximize 
the performance benefits of locality, the hybrid executor followed a simple policy rule:   dispatch tasks to serverless functions only if there are not
local threads available in the pool.

\paragraph{Setup} The maximum dwell of a point was configured to $5$ million iterations.  Each rectangle was subsequently divided in $4$ parts. We run two different configurations: 
the first one with an initial subdivision ($sd$) of $64$ and a maximum recursion depth ($d$) of $5$, and a second one with $sd=256$ and $d=4$. 
Note that the second configuration implies an initial steeper demand for concurrency, leading to an earlier use of serverless functions.
Both parallel and hybrid implementations were executed in a \texttt{c5.12xlarge} EC2 instance (48 vCPUs). We run the parallel version with $48$ concurrent threads.
The hybrid version was  restricted to $24$ local threads to compel the participation of cloud threads.

\begin{figure}[h]
  \centering
  \includegraphics[width=0.9\linewidth]{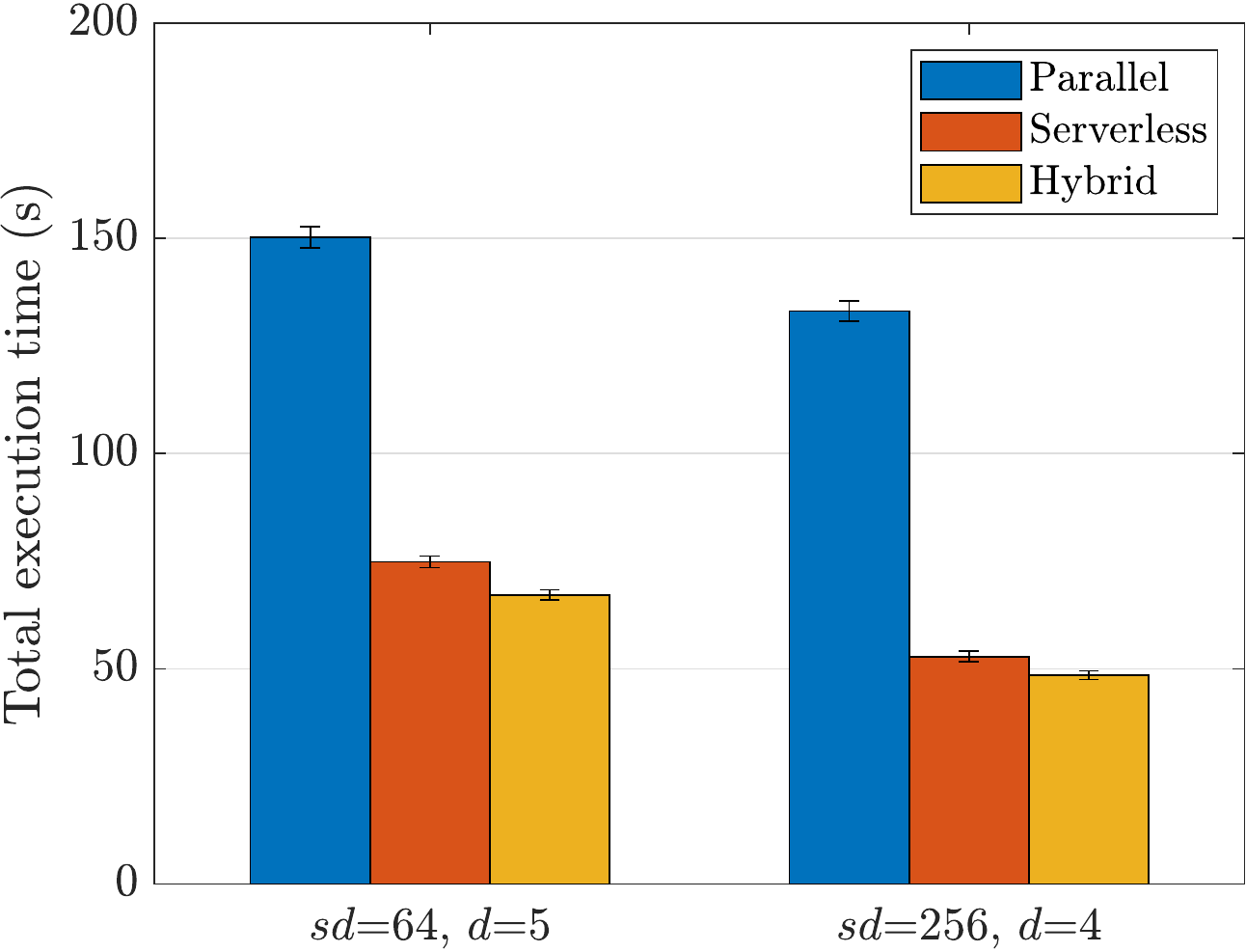}
  \caption{
    Total execution time of Mariani-Silver for different implementations and configurations. The error bars represent the standard deviation of all measurements.
  }
  \label{fig:marianiPerf}
\end{figure}

\paragraph{Results} Figure~\ref{fig:marianiPerf} shows the total execution time of the algorithm under our three different implementations. 
As can be seen in the figure, our serverless and hybrid executors obtain better performance than a large VM, achieving 
a speedup factor $> 2$. Non-surprisingly, the performance of the hybrid executor over the pure serverless executor is not  so high for this
algorithm. The main reason is that the Mariani-Silver algorithm is highly CPU-bound with a very small I/O demand. So, the small improvement
of the hybrid executor  mainly comes from the hiding of the invocation latencies of the remote cloud threads.  We expect significant improvements
for dynamic parallel algorithms where data locality is a key factor for their scalability. Either way, the combination of local and remote cloud resources
makes a case worth exploring in the future.

\subsection{Effects of shared resources}

The Betweenness Centrality algorithm is a CPU-intensive algorithm that also makes a heavy use of data read from memory. In this section we study the performance of this algorithm in a serverless and multithreaded environment. As explained in the section \ref{sec:bc}, and as can be seen in the line \ref{line:bc:functionGraph} of the Listing \ref{lst:bc}, the serverless implementation of the BC algorithm computes in each cloud functions the entire graph. Note that due to the random nature of the graph is impossible to partition it without computing the shortest paths, so to compute the centrality of a node it is necessary to have the entire graph, which is too large to be passed as a parameter to a cloud function. In the multi-threaded implementation, however, the graph can be shared by each task, so there is no need to replicate it. This is the only difference between the serverless and multithreaded versions, the rest of the BC code is exactly the same.

To rule out the effects caused by HyperThreading explained in the \ref{sec:hyperthreading} section, VMs running multithreaded BC code will have HyperThreading disabled for this comparison. These VMs are compared to cloud functions by limiting the maximum number of concurrent functions to the number of CPUs in each VM. In particular, the following AWS specialized compute VMs have been used: \texttt{c5.9xlarge} (18 CPUs), \texttt{c5.12xlarge} (24 CPUs), \texttt{c5.18xlarge} (36 CPUs) and \texttt{c5.24xlarge} (48 CPUs). Client application for the serverless implementation has been executed in a \texttt{c5.2xlarge}.

\begin{figure}[h]
  \centering
  \includegraphics[width=0.9\linewidth]{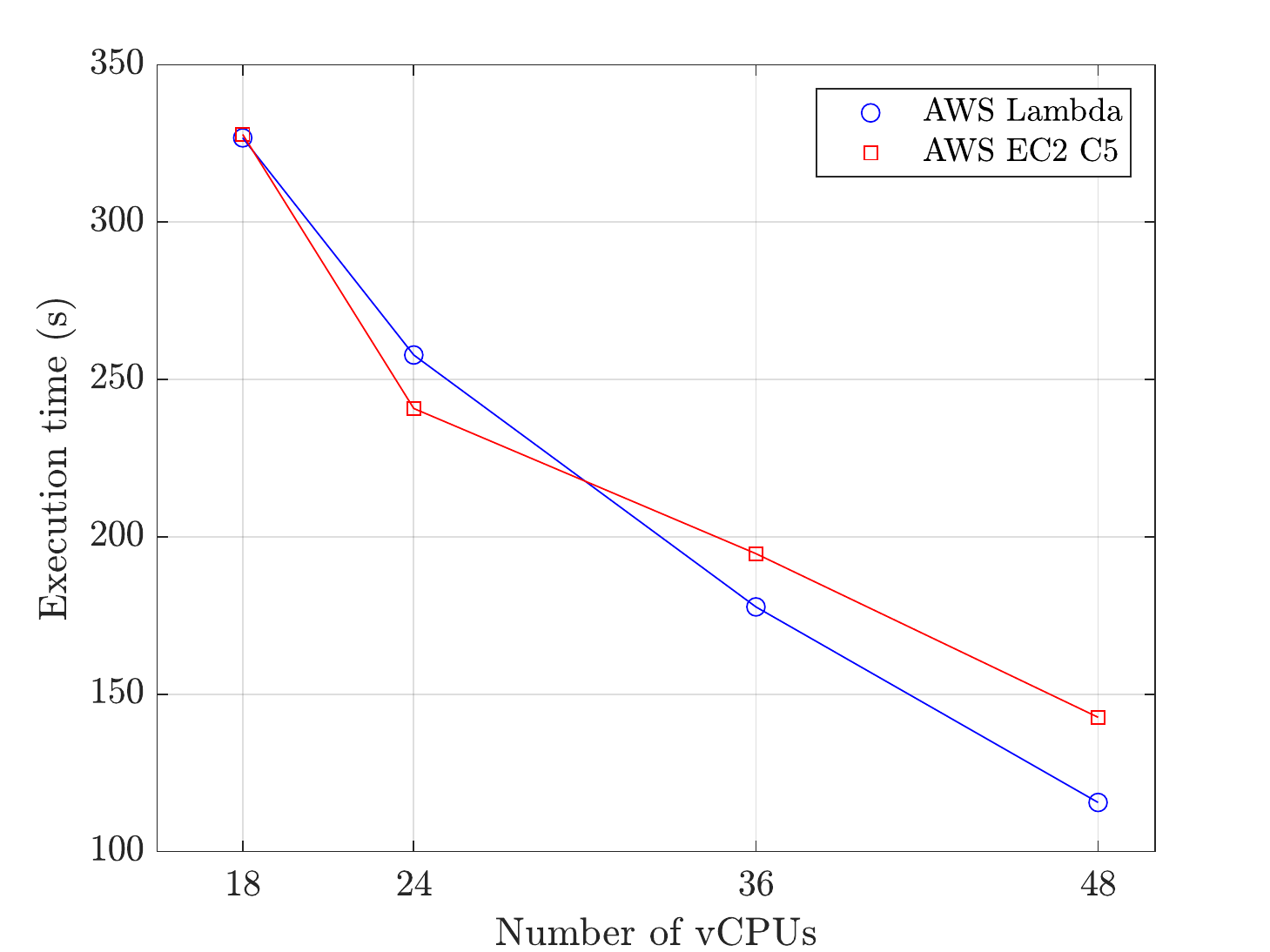}
  \caption{
    Total execution time comparison of Betweenness Centrality ($N=17$) for different implementations.
  }
  \label{fig:bcPlot}
\end{figure}
 
In figure \ref{fig:bcPlot} we show a comparison of the performance of the two implementations. Despite the fact that each serverless function must re-compute the graph and the overheads intrinsic to distributed systems, the serverless version reduces up to $10\%$ the execution time of the parallel version using the same number of vCPUs. In the plot we can also see that the performance of the multithreaded version degrades as the size of the VMs increases. However, the serverless version is able to scale while maintaining performance.

In this experiment it could be assumed that VMs have some advantage. Since they have the same number of CPUs,  Hyperthreading is disabled, and they avoid the overhead of Cloud Function virtualization. However, the serverless implementation is able to perform 10\% better while doing the same exact computations. This is because the performance of the JVM degrades with a high number of threads. Degradation is mainly caused by two factors, the increased Garbage Collector's response time and the usage of shared resources. For this case, shared resources cause drawbacks like memory access contention and inefficient use of the shared L3 cache, which degrade the total performance of the system.
 
\section{Cost analysis}
\label{sec:cost}

Alongside performance, cost is the other main characteristic that must be analyzed in every serverless computing application. 
Although the promise of cost savings (through a true pay-as-you-go model) is one of the features that attract new applications to FaaS, an accurate cost assessment must be carried out when comparing FaaS with other execution environments. 
In fact, several FaaS applications prototypes show a higher cost than comparable traditional IaaS-based applications~\cite{berkeleyServerless}.

Compared to other cloud computing offerings, FaaS price includes much more than the specific resources: scaling, redundancy for availability, monitoring, and logging. 
But if we only take into account the computing power, cheaper alternatives exist also in the cloud. 
Transient servers like Amazon EC2 Spot offer spare compute capacity at steep discounts. 
After recent changes in Amazon Spot pricing model, there are fewer instance interruptions and the prices are more stable and predictable, making transient servers a good option for running exploratory tasks.

A clear scenario where it is worth paying FaaS cost is when the nature of workload causes an underutilization of virtual machine or cluster computing resources. 
It is common to overprovisioning a cluster to cope with peak demand of irregular workloads. 
An alternative, when using cloud computing, is to autoscale resources according to computing demands. 
But startup latency of VM instances is still higher than more lighter serverless containers. Moreover, application programmer have to deal with the complexities of elastic scaling, while serverless offers this out-of-the-box.

Determining the appropriate amount of resources to devote to an application is still a matter of research, specially for irregular workloads. We believe that due to its inherent elasticity, FaaS may be a cost-efficient approach to explore the solution space of a problem, even if cost is a bit higher. 
Serverless functions could be useful to find the right provisioning for a problem before moving to less expensive, but more complex to manage cloud resources, like transient servers.

\begin{figure} [h]
  \centering
  \includegraphics[width=0.9\linewidth]{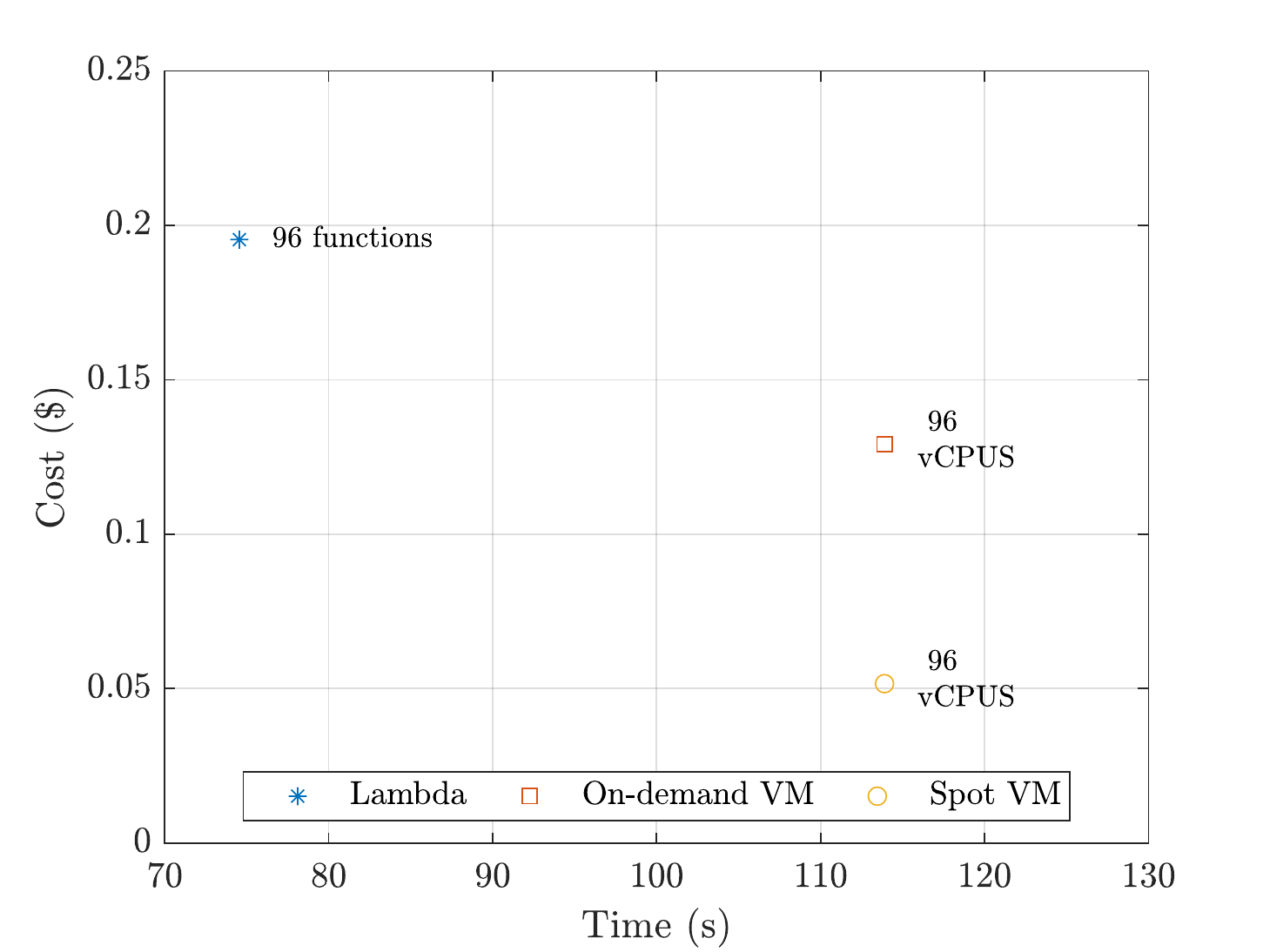}
  \caption{
    Cost-performance comparison of the parallel and serverless version of UTS at depth $17$ with comparable CPU resources.
  }
  \label{fig:utsCostPerformanceLambdaVsVM}
\end{figure}

Figure~\ref{fig:utsCostPerformanceLambdaVsVM} shows the charged costs for the UTS at depth $17$ both in AWS Lambda (with the limit set to $96$ concurrent functions) and using the parallel version in a \texttt{c5.24xlarge} VM. Although the performance of the serverless version is better, the cost is higher, specially if compared to the spot instance pricing model.

\begin{figure} [h]
  \centering
  \includegraphics[width=0.9\linewidth]{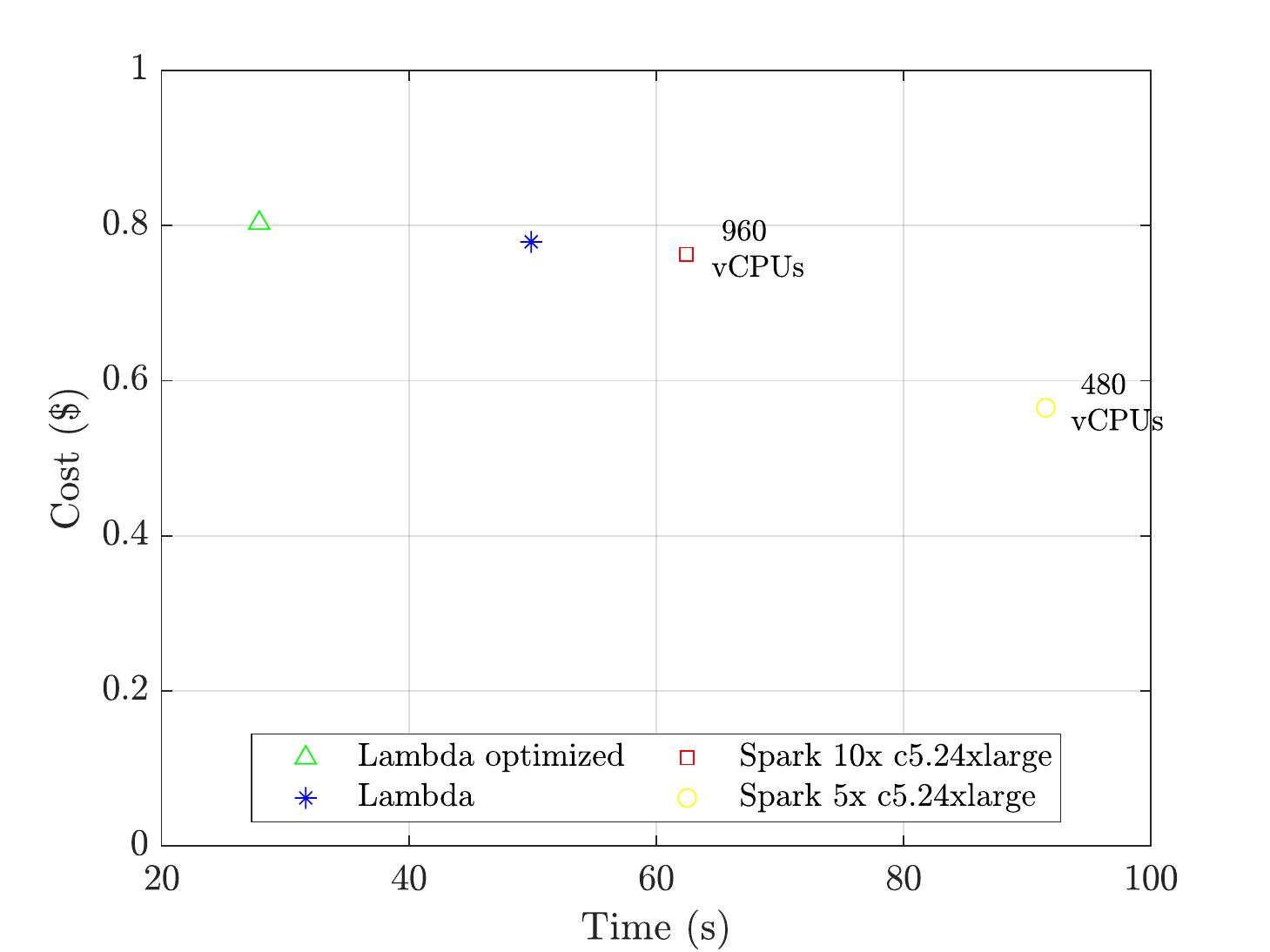}
  \caption{
    Cost-performance comparison of the Spark and serverless version of UTS at depth 18.
  }
  \label{fig:utsCostPerformanceLambdaVsSpark}
\end{figure}

Highly parallelizable algorithms can benefit from the high concurrency and elasticity that FaaS platforms can provide. Figure~\ref{fig:utsCostPerformanceLambdaVsSpark} shows a cost-performance comparison of the serverless version of UTS with a Spark implementation~\cite{resilientx10} that uses a Map/Reduce strategy, with tree traversal divided into rounds. Both versions are parameterized to achieve the best performance possible. The serverless version is limited to $2,000$ concurrent functions, the maximum concurrency supported by the cloud provider in the region we run the experiment. We run the Spark versions with two different settings: an Elastic Map Reduce (EMR) cluster of $10$x \texttt{c5.24xlarge} workers (totalling 960 vCPUs), and a smaller cluster of $5$x \texttt{c5.24xlarge} workers (480 vCPUs). The master node is a smaller \texttt{m5.2xlarge} instance. Costs are calculated on the basis of the on-demand pricing of EMR. 
For example, for the 10-node cluster and considering $t$ is the total execution time in seconds:

\begin{equation}
Cost_{EMR}= \frac{\textrm{t}}{3600} (10 \cdot 4.35\textrm{\$ per hour} + 0.48\textrm{\$ per hour})
\label{eq:costEMR}
\end{equation}

We see that the unoptimized serverless version depicted in Figure~\ref{fig:utsCostPerformanceLambdaVsSpark} is able to outperform Spark by up to $20\%$, for a similar cost. If we calculate the price to performance ratio, dividing the performance in million nodes per second by the cost in dollars, we find that the serverless version has a ratio of $6,966$ M nodes/s/\$, while the Spark version only achieves $5,689$ M nodes/s/\$. If we apply the dynamic optimizations described in section~\ref{sec:performance}, the serverless version is able to outperform Spark by up to $55\%$. With the dynamic optimizations the price to performance ratio achieved is  $11,366$ M nodes/s/\$ thanks to the great time reduction and the low cost increase.

\begin{figure}[h]
  \centering
  \includegraphics[width=0.9\linewidth]{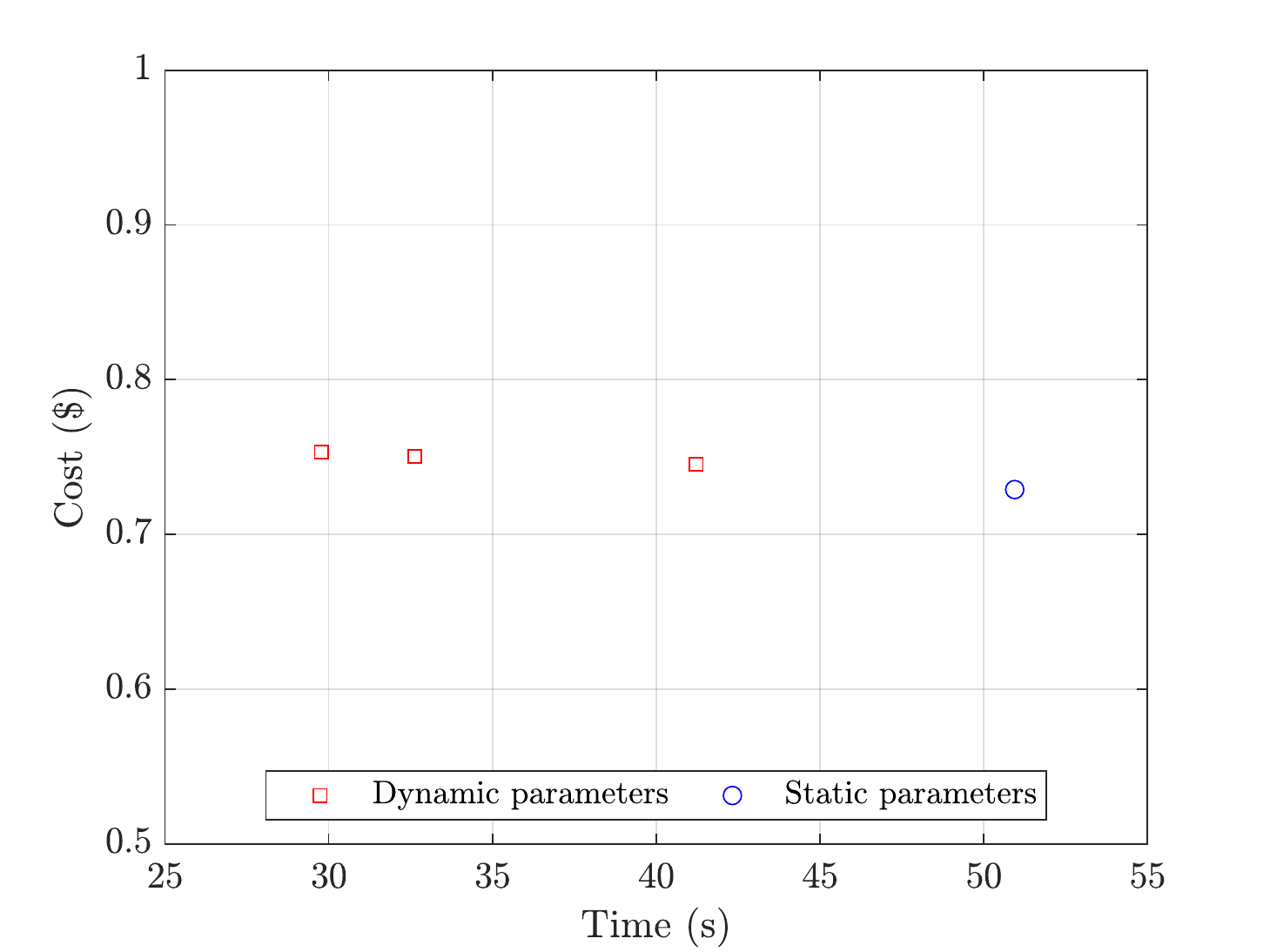}
  \caption{
    Cost-performance comparison of UTS at depth 18 using static and dynamic parameters.
  }
  \label{fig:utsCostPerformanceLambdaStaticVsDynamic}
\end{figure}

In fact, the cost of serverless executions hardly depends on the total execution time. In the case of the UTS, the main challenge is to be able to have the maximum number of workers exploring the tree, so that the average amount of work they perform is large enough to compensate the system overhead. As explained in the previous section, the optimization of dynamic parameters can be used during the exploration. At times when you have fewer workers, you can try to divide the tasks more and when you have the pool of workers at full capacity, divide them into fewer tasks. Similarly, when you have few workers it is interesting that workers don't explore too much in order to create new tasks soon, instead when you have many workers you can allow them to explore more in order to try to reduce overheads.

With this in mind, if we look at Eq. \ref{eq:costServerless} that makes explicit the cost of serverless executions, the sums of invocations and the cost of the VM are orders of magnitude below the price of serverless functions' execution time. As the computation time used by serverless functions is practically constant and thanks to the pay-as-you-go billing model, the most aggressive optimization strategies have approximately the same cost as the execution with static parameters. This fact allows, to the UTS optimizations, to achieve an improvement in performance at a very low extra cost.

Figure \ref{fig:utsCostPerformanceLambdaStaticVsDynamic} shows the cost and wall time of different executions of the UTS with different strategies of dynamic parameters together with the metrics of the static execution. It can be seen that the cost difference does not exceed $0.024\$$, which represents $3.31\%$ of the total cost of the static execution. However, an execution time improvement of $41.56\%$ is achieved compared to the with the static parameters.


In section~\ref{sec:performance} we already saw that our serverless and hybrid executors can obtain better performance than a multithreaded executor in a large virtual machine for the Mariani-Silver algorithm. In Table~\ref{tab:mariani}, we show the costs of the Mariani-Silver with the configurations $sd$=256 and $d$=4. VM costs are accounted for according to on-demand instance pricing without considering VM initialization times. The minimum billing period for VM is $1$s. Although the cost of the parallel multithreaded version is lower than the versions that use serverless functions, our hybrid serverless executor is the most cost-effective implementation, achieving the highest price to performance ratio.

\begin{table}
  \caption{
    Cost/Performance of Mariani-Silver serverless and parallel implementations.
  }
  \label{tab:mariani}
  \begin{tabular}{lllp{2.1cm}}
    & Time(s) & Cost (\$) & Price-Performance\newline Ratio (MP/s/\$) \\ \hline

		Parallel (c5.12xlarge) &  $133.13$ & $0.0852$ & $1.47$   \\ \hline
		Serverless &  $52.85$ & $0.2360$ & $1.34$  \\ \hline
		Hybrid &  $48.53$ & $0.2041$ & $1.69$   \\ \hline

		\bottomrule
  \end{tabular}
\end{table}

\section{Conclusions}

\label{sec:conclusion}
In this work, we have validated the hypothesis that the high elasticity of the FaaS execution model can be  key enabler for effectively running dynamic and irregular task-parallel applications in the cloud. Through an evaluation focused on performance and cost, we see that FaaS can achieve a good performance for high-concurrency algorithms like UTS, thus masking the latency of invoking remote functions. We demonstrate better performance than a Spark cluster for similar cost. Further, we prove that a hybrid implementation combining local threads and serverless functions can be more cost-efficient than a pure serverless implementation for recursive algorithms that demand high levels of concurrency. Finally, we show that a pure serverless solution can outperform a large EC2 VM in the Betweeness Centrality workload when using the same amount of virtual CPUs. Consequently, we are safe to conclude that a simple executor similar to those available in the Java Concurrency API can be sufficient to run high concurrency algorithms on serverless platforms, thus paving the way towards a more transparent and elastic execution of those problematic algorithms at large scale.

In the future, we would like to extend and further validate our work with other algorithms that involve stateful computations, like complex graph algorithms, that usually operate on I/O-bound irregular and unbalanced workloads. We also plan to study optimal hybrid deployment strategies to adapt executors to different application workloads.

\section*{Acknowledgements}

This work has been partially supported by the EU Horizon 2020 programme under grant agreement No 825184,  
and by the Spanish Government through project PID2019-106774RB-C22. 
Marc Sánchez-Artigas is a Serra Húnter Fellow.


\bibliography{mybibfile}

\end{document}